\pdfoutput=1

\documentclass[sigconf]{acmart}

\usepackage{booktabs} 

\usepackage{array}
\newcommand{\PreserveBackslash}[1]{\let\temp=\\#1\let\\=\temp}
\newcolumntype{C}[1]{>{\PreserveBackslash\centering}p{#1}}
\newcolumntype{R}[1]{>{\PreserveBackslash\raggedleft}p{#1}}
\newcolumntype{L}[1]{>{\PreserveBackslash\raggedright}p{#1}}

\usepackage{amsmath}
\usepackage{bm}
\usepackage{amsfonts}
\usepackage{amsthm}
\usepackage{multirow}
\usepackage{colortbl}
\usepackage{enumitem}
\usepackage{subcaption}
\usepackage{tabu}
\usepackage{balance}
\usepackage{footnote}
\makesavenoteenv{tabular}
\makesavenoteenv{table}
\usepackage{mathtools}

\usepackage{fontawesome}

\def\eqref#1{equation~\ref{#1}}

\def\1{\bm{1}}

\DeclareMathAlphabet{\mathsfit}{\encodingdefault}{\sfdefault}{m}{sl}
\SetMathAlphabet{\mathsfit}{bold}{\encodingdefault}{\sfdefault}{bx}{n}

\DeclareMathOperator*{\argmin}{arg\,min}

\DeclareMathOperator*{\bertcat}{BERT_\text{CAT}}
\DeclareMathOperator*{\bertdot}{BERT_\text{DOT}}
\DeclareMathOperator*{\bert}{BERT}

\DeclareMathOperator*{\colbert}{ColBERT}

\usepackage{array, booktabs}
\usepackage{graphicx}

\definecolor{niceBlue}{RGB}{66, 109, 179}
\definecolor{niceRed}{RGB}{186, 56, 65}

\usepackage{pifont}
\usepackage[noabbrev]{cleveref}

\newcommand{\tsc}[1]{\textsuperscript{#1}} 
\author{Sebastian Hofst{\"a}tter\tsc{1}, Sheng-Chieh Lin\tsc{2}, Jheng-Hong Yang\tsc{2}, Jimmy Lin\tsc{2}, Allan Hanbury\tsc{1}}
\affiliation{
  \institution{\vskip .1cm}
  \institution{\tsc{1} TU Wien, \tsc{2} University of Waterloo}
}

\renewcommand{\authors}{Sebastian Hofst{\"a}tter, Sheng-Chieh Lin, Jheng-Hong Yang, Jimmy Lin, Allan Hanbury}

\copyrightyear{2021} 
\acmYear{2021} 
\setcopyright{acmlicensed}\acmConference[SIGIR '21]{Proceedings of the 44th International ACM SIGIR Conference on Research and Development in Information Retrieval}{July 11--15, 2021}{Virtual Event, Canada}
\acmBooktitle{Proceedings of the 44th International ACM SIGIR Conference on Research and Development in Information Retrieval (SIGIR '21), July 11--15, 2021, Virtual Event, Canada}
\acmPrice{15.00}
\acmDOI{10.1145/3404835.3462891}
\acmISBN{978-1-4503-8037-9/21/07}

\begin{document}

\title{Efficiently Teaching an Effective Dense Retriever with~Balanced~Topic~Aware~Sampling}

\begin{abstract}

A vital step towards the widespread adoption of neural retrieval models is their resource efficiency throughout the training, indexing and query workflows. The neural IR community made great advancements in training effective dual-encoder dense retrieval (DR) models recently. A dense text retrieval model uses a single vector representation per query and passage to score a match, which enables low-latency first-stage retrieval with a nearest neighbor search. Increasingly common, training approaches require enormous compute power, as they either conduct negative passage sampling out of a continuously updating refreshing index or require very large batch sizes. Instead of relying on more compute capability, we introduce an efficient topic-aware query and balanced margin sampling technique, called TAS-Balanced. We cluster queries once before training and sample queries out of a cluster per batch. We train our lightweight 6-layer DR model with a novel dual-teacher supervision that combines pairwise and in-batch negative teachers. Our method is trainable on a single consumer-grade GPU in under 48 hours. We show that our TAS-Balanced training method achieves state-of-the-art low-latency (64ms per query) results on two TREC Deep Learning Track query sets. Evaluated on NDCG@10, we outperform BM25 by 44\%, a plainly trained DR by 19\%, docT5query by 11\%, and the previous best DR model by 5\%. Additionally, TAS-Balanced produces the first dense retriever that outperforms every other method on recall at any cutoff on TREC-DL and allows more resource intensive re-ranking models to operate on fewer passages to improve results further.

\end{abstract}

\maketitle

\begin{figure}[t]
   \includegraphics[trim={.8cm 1cm .5cm 0cm},clip,width=0.43\textwidth]{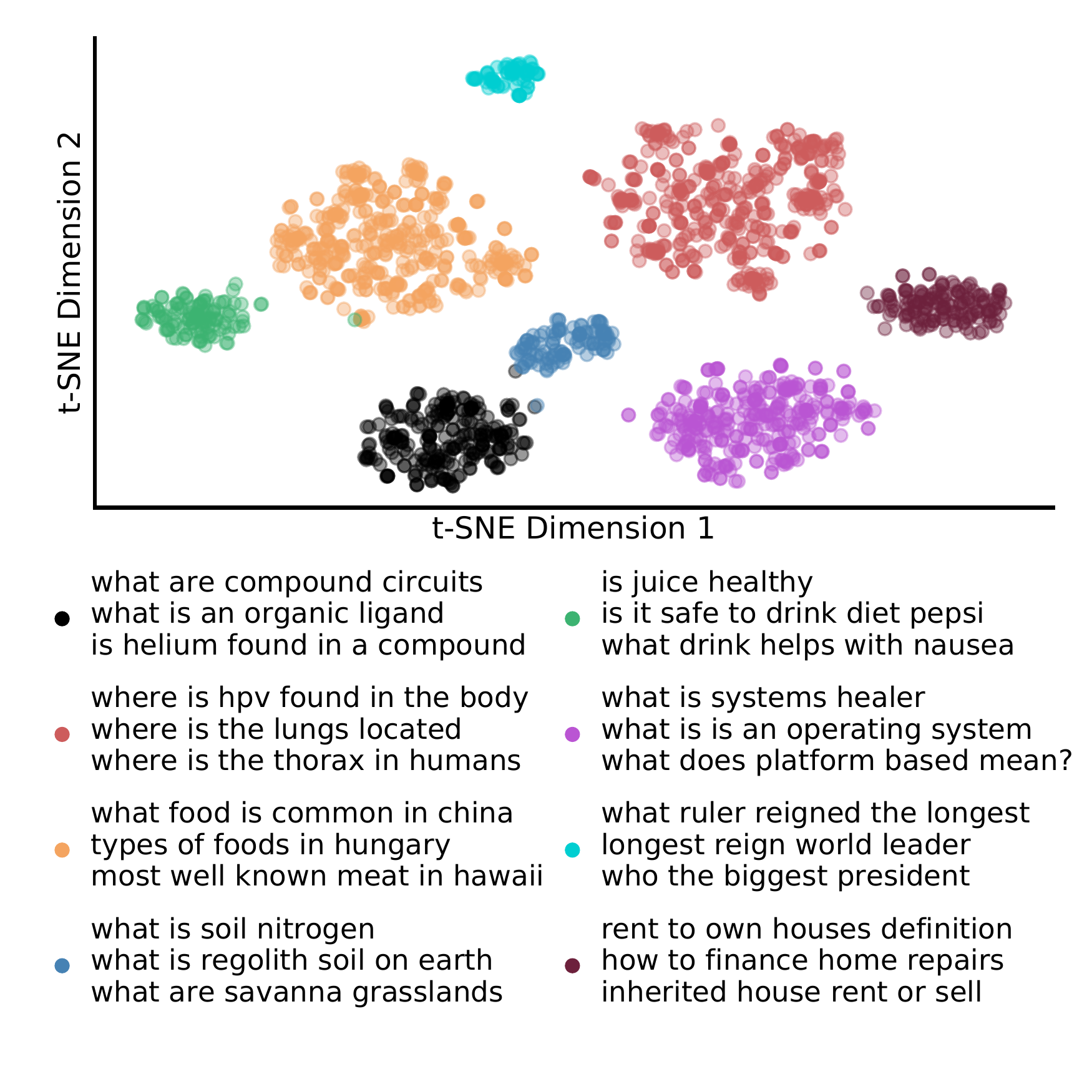}
    \centering
    \vspace{-0.3cm}
    \caption{T-SNE plot of 8 randomly sampled topic clusters and example queries. Our Topic Aware Sampling (TAS) composes queries from a single cluster per batch.}
    \label{fig:topic_hero}
    \vspace{-0.4cm}
\end{figure}

\section{Introduction}

Having a well prepared teacher in life makes learning easier and more efficient. Training dense text retrieval models with more experienced and capable teacher models follows the same path. Dense retrieval models -- such as the BERT-based \cite{devlin2018bert} dual-encoder $\bertdot$ -- offer the great potential of low-latency query times, vastly better accuracy and recall than traditional first-stage retrieval methods, and moving most computational cost into offline indexing and training. The unifying $\bertdot$ architecture is already supported by many open source search engines. $\bertdot$ can be used as a standalone retriever or as part of a re-ranking pipeline. The problem, when further improving the result quality, becomes the affordability in terms of hardware resources and requirements for training and indexing. A recent trend to improve retrieval result quality is to augment the $\bertdot$ training procedure, which leads to increased hardware requirements. Examples of this include conducting negative passage sampling out of a continuously updating refreshing index (ANCE \cite{xiong2020approximate}), generations of models (LTRe \cite{zhan2020learning}), or requiring large batch sizes (RocketQA \cite{ding2020rocketqa}). 

A concurrent line of inquiry is the use of knowledge distillation from more effective, but less efficient architectures as teachers either in pairwise \cite{hofstaetter2020_crossarchitecture_kd,lu2020twinbert,izacard2020distilling} or in-batch negatives \cite{lin2020distilling} settings. In-batch negatives reuse the encoded representations per sample and compute interactions between all samples in a batch. We combine these two knowledge distillation paradigms into a novel dual-supervision with a pairwise concatenated $\bertcat$ and a $\colbert$ teacher for in-batch negatives. These approaches, while already working well, are constrained by the information gain a single random batch can deliver for training. The training data available to dense retrieval training consists of a pool of queries, and associated with each query is typically a set of passage pairs with a teacher score margin.
Each pair consists of a relevant and non-relevant passage, with the margin set by subtracting the non-relevant sampled passage teacher score from the relevant passage teacher score.

The main contribution of this work is to improve both pairwise and in-batch teacher signals. We propose Balanced Topic Aware Sampling (TAS-Balanced) to compose dense retrieval training batches. This sampling technique has two components: (1) we compose batches based on queries clustered in topics (TAS); and (2) we then select passage pairs so as to balance pairwise teacher score margins (TAS-Balanced). 
We cluster the topics once before training based on a baseline representation by semantic dot product similarity (which allows grouping queries without lexical overlap) -- a one time cost of under 10 minutes for all 400K training queries of MSMARCO. An example selection of topic clusters is shown in Figure \ref{fig:topic_hero}. 
Previously, a batch would be composed of random queries from the training set, leaving little information gain for in-batch negatives. By selecting queries from a single cluster, we concentrate information about a topic in a single batch, which after in-batch negative teaching, leads to higher quality retrieval results.  

We show that with TAS-Balanced batches and dual-supervision we can train a very effective dense retrieval model on a single consumer-grade (11GB memory) GPU in under 48 hours, as opposed to a common configuration of 8x V100s, because our method does not rely on repeated indexing \cite{xiong2020approximate} or large batch size training \cite{ding2020rocketqa}. Specifically, we study the following research questions:

\newcommand{\RQone}{\begin{itemize}
    \item[\textbf{RQ1}] How effective are TAS and TAS-Balanced batch sampling techniques with single and dual-teacher supervision?
\end{itemize}}
\newcommand{\RQoneRunning}{\textbf{RQ1} \textit{..? }}
\RQone

\noindent We find TAS improving both in-batch negative teaching alone as well as our dual-supervision teachers. The TAS-Balanced sampling improves pairwise training, in-batch negatives, and the dual-supervision training, which represents the best overall configuration across our three query sets. The dual-teacher supervision has an especially big positive impact on recall using a Margin-MSE loss. We study different losses for the dual-supervision and find  Margin-MSE improves the results consistently over other loss functions. 

A common problem in machine learning research is inadvertent overfitting on a specific combination of hyperparameters, random seed, and collection. To gain confidence in our results, we study:

\newcommand{\RQtwo}{\begin{itemize}
    \item[\textbf{RQ2}] How robust is TAS-Balanced to different randomization?
\end{itemize}}
\newcommand{\RQtwoRunning}{\textbf{RQ2} \textit{How robust is TAS-Balanced to different randomization?}}
\RQtwo

\noindent To show that TAS-Balanced is robust against random overfitting, we conduct a randomization study of 5 instances with different random orderings of selected clusters and queries. We find only small standard deviations across the metrics of our query sets ($<.01$ nDCG change on TREC-DL; $<.001$ MRR on MSMARCO-DEV). This gives us great confidence in the efficacy and robustness of our technique.
To set our results in context to related work, we answer:

\newcommand{\RQthree}{\begin{itemize}
    \item[\textbf{RQ3}] How does our TAS-Balanced approach compare to other dense retrieval training methods?
\end{itemize}}
\RQthree
We evaluate our models on two TREC-DL ('19 \& '20) query sets and the MSMARCO-DEV set using the MSMARCO passage collection. The two TREC sets are especially suited to study recall quality of the dense retrievers with hundreds of judged passages per query. Our TAS-Balanced \& dual-supervision trained $\bertdot$ model shows state-of-the-art low-latency results on both TREC-DL'19 and '20 query sets using a batch size as small as 32. Our $\bertdot$ model, evaluated on NDCG@10, outperforms BM25 by 44\%, a plainly trained DR by 19\%, docT5query by 11\%, and the previous best DR model by 5\%. On the sparse labelled MSMARCO-DEV queries, TAS-Balanced shows the best results for methods using a single consumer-grade GPU and outperforms most approaches that require 20x more resources to train. 
Finally, while TAS-Balanced is an effective standalone low-latency retriever, we also study the impact of our TAS-trained model in a larger search system:

\newcommand{\RQfour}{\begin{itemize}
    \item[\textbf{RQ4}] How well suited is our TAS-trained dense retriever as a first-stage module in terms of recall and re-ranking gains?
\end{itemize}}
\RQfour

\noindent We find that TAS-Balanced results in the first $\bertdot$ model that outperforms BM25 and docT5query consistently on every recall cutoff in TREC-DL densely judged query sets. Fused together with docT5query results we see another increase in recall, showing that dense and sparse solutions still benefit from each other at an already high recall level. Furthermore, we stack the state-of-the-art re-ranking system mono-duo-T5 on top of our first-stage retrieval. Because TAS-trained $\bertdot$ increases the recall and accuracy for small cutoffs, we can reduce the number of passages an expensive re-ranking system processes and still receive considerable benefits. However, we also find a limitation in re-ranking quality for higher cutoffs: Even though TAS-Balanced continues to improve the recall at higher cutoffs, the re-ranking does not take advantage of that. Future work may improve re-rankers on top of TAS-Balanced. 

The aim of this work is to produce a very effective $\bertdot$ retrieval model and minimize the training resources necessary. Our contributions are as follows:
\begin{itemize}[leftmargin=*]
    \item We propose an efficient Topic Aware Sampling (TAS-Balanced) for composing informative dense retrieval training batches
    \item We show that TAS-Balanced in combination with a dual-teacher supervision achieves state-of-the-art DR results on TREC-DL
    \item We study our training robustness and how TAS-Balanced improves a larger (re-)ranking system
    \item We publish our source code at: \\ {\small\url{https://github.com/sebastian-hofstaetter/tas-balanced-dense-retrieval}} %
\end{itemize}

\vspace{-0.2cm}
\section{Retrieval Model Background}
\label{sec:models}

We employ three different Transformer based \cite{vaswani2017attention} \& BERT pre-trained \cite{devlin2018bert} architectures in our work. We use two teacher architectures for the best combination of pairwise ($\bertcat$) and in-batch negative teaching ($\colbert$) to train our our main dense retrieval model: the dual-encoder BERT$_\text{DOT}$ architecture. In the following we present the characteristics of each model architecture, our dual-teacher supervision, as well as related training methods.

\subsection{BERT Teacher Models}

The common way of utilizing the BERT pre-trained Transformer model in a re-ranking scenario is by concatenating query and passage input sequences \cite{nogueira2019passage,macavaney2019,yilmaz2019cross}. We refer to the architecture as BERT$_\text{CAT}$. We use it in this work as our strong pairwise teacher model. In the BERT$_\text{CAT}$ ranking model, the query ${q}_{1:m}$ and passage ${p}_{1:n}$ sequences are concatenated with special tokens (using the $;$ operator), encoded with BERT, the CLS token representation pooled, and scored with single linear layer $W_s$:
\begin{equation}
\begin{aligned}
    \bertcat({q}_{1:m},{p}_{1:n}) & = W_s\bert(\text{CLS};{q}_{1:m};\text{SEP};{p}_{1:n})_\text{CLS}
\end{aligned}
\end{equation}
\noindent This architecture is easy to train and provides very strong results in terms of effectiveness, especially when used in an ensemble \cite{hofstaetter2020_crossarchitecture_kd}.
However, it requires candidate selection prior to re-ranking, has no ability to pre-compute and index passage representations, and is therefore slow in practice \cite{Hofstaetter2019_osirrc}. 

The $\colbert$ model \cite{khattab2020colbert} tries to overcome the time-efficiency problem of $\bertcat$ by delaying the interactions between every query and document term representation after BERT encoding:
\begin{equation}
\begin{aligned} 
\hat{q}_{1:m} &= \bert(\text{CLS};{q}_{1:m};\text{SEP}) \\
\hat{p}_{1:n} &= \bert(\text{CLS};{p}_{1:n};\text{SEP})
\end{aligned}
\end{equation}
The interactions in the $\colbert$ model are aggregated with a max-pooling per query term and sum of query-term scores as follows:
\begin{equation}
\begin{aligned}
    \colbert({q}_{1:m},{p}_{1:n}) = \sum_{1}^{m} \max_{1..n} \hat{q}_{1:m}^T \cdot \hat{p}_{1:n}
\end{aligned}
\end{equation}
This decoupling of query and passage encoding allows the passage representations to be indexed in theory.
However, the storage cost of pre-computing passage representations is much higher and scales in the total number of terms in the collection. Because of the storage increase and increased complexity for the scoring aggregation we refrain from using $\colbert$ as a dense retrieval model, and rather use it as an efficient teacher for in-batch negatives.

\subsection{BERT\texorpdfstring{$_\textbf{DOT} $} : Dense Retrieval Model}

The BERT$_\text{DOT}$ model encodes query ${q}_{1:m}$ and passage ${p}_{1:n}$ sequences independently from each other and matches only a single representation vector of the query with a single representation vector of a passage \cite{xiong2020approximate,luan2020sparse,lu2020twinbert}. 
It pools each CLS token output for query $\hat{q}$ and passage $\hat{p}$ representations as follows:
\begin{equation}
\begin{aligned} 
\hat{q} = \bert(\text{CLS};{q}_{1:m};\text{SEP})_\text{CLS} \ , \
\hat{p} = \bert(\text{CLS};{p}_{1:n};\text{SEP})_\text{CLS}
\end{aligned}
\end{equation}
Potentially after storing all representations in an index, the model computes the final scores as the dot product $\cdot$ of $\hat{q}$ and $\hat{p}$:
\begin{equation}
\begin{aligned}
    \bertdot({q}_{1:m},{p}_{1:n}) & = \hat{q} \cdot \hat{p}
\end{aligned}
\end{equation}
\noindent The independence of query and document encoding as well as the dot product scoring enables two crucial operations for this work. First, we encode all queries once and use their representation for clustering in our TAS approach and second we deploy BERT$_\text{DOT}$ with a simple maximum-inner product retrieval workflow: After training, we encode and index every passage once in a nearest neighbor search index and retrieve the top results at query time for a single encoded query.

\begin{table}[t]
    \centering
    \caption{Latency analysis of Top-1000 retrieval using our $\bertdot$ retrieval setup for all MSMARCO passages using DistilBERT and Faiss (FlatIP) on a single Titan RTX GPU}
    \label{tab:latency_analysis}
    \setlength\tabcolsep{2.4pt}
    \vspace{-0.3cm}
    \begin{tabular}{l!{\color{lightgray}\vrule}rr!{\color{lightgray}\vrule}rr!{\color{lightgray}\vrule}rr}
       
       \toprule

       \textbf{Batch} &
       \multicolumn{2}{c!{\color{lightgray}\vrule}}{\textbf{Q. Encoding}}&
       \multicolumn{2}{c!{\color{lightgray}\vrule}}{\textbf{Faiss Retrieval}}&
       \multicolumn{2}{c}{\textbf{Total}}\\
       \textbf{Size}&
       Avg.&
       $99^\text{th}$ Per.&
       Avg.&
       $99^\text{th}$ Per.&
       Avg.&
       $99^\text{th}$ Per.\\

       \midrule
       \arrayrulecolor{lightgray}
        \textbf{1}     & 8 ms & 11 ms & 54 ms  & 55 ms  & 64 ms  & 68 ms \\ 
        \textbf{10}    & 8 ms & 9 ms  & 141 ms & 144 ms & 162 ms & 176 ms   \\ 
        \textbf{2,000} & 273 ms & 329 ms & 2,515 ms & 2,524 ms & 4,780 ms & 4,877 ms  \\ 
        \arrayrulecolor{black}
       \bottomrule
    \end{tabular}
\end{table}

In Table \ref{tab:latency_analysis} we give a training-agnostic latency analysis of our $\bertdot$ retrieval setup. We use both the DistilBERT encoder and a brute-force Faiss nearest neighbor index (FlatIP) on a single TITAN RTX GPU with a total of 24 GB memory. We can fit a batch size of up to 2,000 queries on this single GPU. We measure that a single query can be responded to in 64ms, batching up to 10 queries (for example in a high load system) only reduces the latency to 162ms. The practical effect of always computing roughly the same number of operations reduces the volatility of the latency to a very small margin as the 99$^\text{th}$ percentile of the measured latency is very close to the mean. For our one-time clustering we utilize the query encoding only with a batch size of 2,000, shown in Table \ref{tab:latency_analysis}. The fast processing allows us to encode all 400K MSMARCO training queries in one minute.

\begin{table*}[t!]
    \centering
    \caption{Comparison of the computational cost of dense retrieval training methods. The GPUs refer to classes: V100 stands for a server-grade GPU with $\geq$ 32 GB memory; GTX refers to a consumer-grade GPU $\geq$ 11GB memory (GTX 1080Ti or better).}
    \label{tab:related_overview}
    \vspace{-0.3cm}
    \setlength\tabcolsep{3pt}
    \begin{tabular}{llll!{\color{lightgray}\vrule}ll!{\color{lightgray}\vrule}ll!{\color{lightgray}\vrule}l}
       \toprule
       &\multirow{2}{*}{\textbf{Training}} & \multirow{2}{*}{\textbf{Min. GPU}} & \textbf{Batch} & \multirow{2}{*}{\textbf{KD Teacher}} & \textbf{Added Cost} & {\footnotesize\faRepeat } \textbf{Passage} & \textbf{Index} & \multirow{2}{*}{\textbf{Misc. Costs}} \\
       
       && & \textbf{Size} && (per Sample) & \textbf{Sampling} & \textbf{Refresh} \\

        \midrule
        &\textbf{Standalone}                                        & 1$\times$ GTX & 32 & -- & -- & -- & -- & -- \\
        \cite{xiong2020approximate} &\textbf{ANCE}                            & 8$\times$ V100 & 32 & -- & -- & \checkmark & 10K batches & +1 BM25-trained checkpoint \\
        \cite{zhan2020learning}&\textbf{LTRe}                                & 1$\times$ GTX  & 32 & -- & -- & \checkmark & 1$\times$ & +1 ANCE checkpoint \\
        \cite{hofstaetter2020_crossarchitecture_kd}&\textbf{Margin-MSE}      & 1$\times$ GTX  & 32 & BERT$_\text{CAT}$ & $\times$ 1-3 & -- & -- & -- \\
        \cite{lin2020distilling}&\textbf{TCT}                                & 1$\times$ V100 & 96 & $\colbert$ & $\times$ 1 & -- & -- & +1 BM25-trained checkpoint \\
        \cite{ding2020rocketqa}&\textbf{RocketQA}                           & 8$\times$ V100 & 4,000 & BERT$_\text{CAT}$ & $\times$ > 13 & \checkmark & 2$\times$ & 4$\times$ cycles of training\\ 
        \midrule
        &\textbf{TAS-Balanced}                                                        & 1$\times$ GTX  & 32 & BERT$_\text{CAT}$ + $\colbert$ & $\times$ 1-4 & -- &  -- & 1$\times$ query clustering \\
        
        \bottomrule
    \end{tabular}
    \vspace{-0.3cm}
\end{table*}

\subsection{Dual-Teacher Supervision}

The community produces mounting evidence that knowledge distillation is essential for effective dense retrieval training \cite{hofstaetter2020_crossarchitecture_kd,lin2020distilling,ding2020rocketqa}. \citet{hofstaetter2020_crossarchitecture_kd} showed the benefits of an ensemble of pairwise $\bertcat$ teachers; concurrently,  \citet{lin2020distilling} showed the benefits of using a $\colbert$ teacher model for in-batch negative sampling. Both possess unique strengths: $\bertcat$ is the more effective teacher, but prohibitively expensive to use for in-batch negatives as it requires quadratic scaling in the batch size, because we need to encode concatenated pairs; $\colbert$ only requires a linear runtime (in the batch size) for in-batch negative scores.

In this work we combine these two approaches into a novel dual-teacher supervision paradigm that provides the best trade-off between effective teaching and efficient training. 

We utilize the published $\bertcat$ ensemble scores from \citet{hofstaetter2020_crossarchitecture_kd} for every official training triple of the MSMARCO-Passage collection. Using this data allows us to use these teacher model $M_t$ scores as a teacher signal for our $M_s$ student model ($\bertdot$) without computational cost. Any pairwise loss function is applicable here, we use the very effective Margin-MSE loss \cite{hofstaetter2020_crossarchitecture_kd}, formalized as follows:
\begin{equation}
\begin{aligned} 
\mathcal{L}_{Pair}(Q,P^{+},P^{-}) = \operatorname{MSE}(&M_s(Q,P^{+}) - M_s(Q,P^{-}),\\ &M_t(Q,P^{+}) - M_t(Q,P^{-}))
\end{aligned}
\end{equation}
For the in-batch negative signal, we use the fact that both $\bertdot$ student and $\colbert$ teacher can independently compute the representation vectors, that is then scored via a dot-product. To create in-batch pairings we cross the representation and pair each positive passage with all other passages in the batch and compute the loss:
\begin{equation}
\begin{aligned} 
\mathcal{L}_{InB}(Q,P^{+},P^{-}) = \frac{1}{2|Q|} &\big( \sum_i^{|Q|} \sum_{p^-}^{P^{-}} \mathcal{L}_{Pair}(Q_i,P_i^+,p^-) \\&+ \sum_i^{|Q|} \sum_{p^+}^{P^{+}} \mathcal{L}_{Pair}(Q_i,P_i^+,p^+) \big)
\end{aligned}
\end{equation}
Here, for simplicity and harmony between the dual-teacher signals we re-use the $\mathcal{L}_{Pair}$ loss, but the teacher model $M_t$ is now $\colbert$. Additionally, we studied list-based losses that score each query with the full set of all passages in the batch, as shown in Section \ref{sec:ablations}, and found Margin-MSE to be the best choice. We compute the total loss as the weighted combination  (where $\alpha$ is a hyperparameter to steer the influence) of the pairwise loss $\mathcal{L}_{Pair}$ and the in-batch loss $\mathcal{L}_{InB}$ as follows:
\begin{equation}
\begin{aligned} 
\mathcal{L}_{DS}(Q,P^{+},P^{-}) = \mathcal{L}_{Pair}(Q,P^{+},P^{-}) + \mathcal{L}_{InB}(Q,P^{+},P^{-}) \times \alpha 
\end{aligned}
\end{equation}

\noindent Following, the findings of \citet{hofstaetter2020_crossarchitecture_kd} and \citet{ding2020rocketqa} we do not use the binary relevance labels provided by MSMARCO directly and only rely on the teacher supervision signal, which confirms in almost all cases the binary relevance ordering. This allows our method to be applied to unsupervised scenarios, where training data is generated and scored by trained teacher models without human assessments. 

\subsection{Other Dense Retrieval Training Methods}

%
%
Improving the training of the $\bertdot$ model is a rapidly evolving field in neural IR with a variety of approaches with different training costs. 
To give a structured overview of the state of the field, we summarize and compare the most related work for dense passage retrieval training with our TAS-Balanced approach in Table \ref{tab:related_overview}. We identify three main drivers of computational cost per training method that lead to a minimum GPU requirement per method. First, the recommended batch size; second, whether knowledge distillation is used; and third, if a dynamic index refresh is needed during training. The standalone training of the $\bertdot$ model only uses binary relevance labels and BM25-sampled negative passages \cite{karpukhin2020dense}. While it offers the lowest cost training, its results are inferior to the other methods, as we show in Table \ref{tab:tas_teacher_abl}.

The ANCE \cite{xiong2020approximate} training swapped BM25-generated negative samples for negative samples retrieved from an index that needs to be refreshed fully every 10K batches, which according to \citet{xiong2020approximate} requires 8 GPU-hours every 10K batches for MSMARCO. \citet{zhan2020learning} built upon ANCE with LTRe training by continuing to train the query encoder with a fixed passage encoder module. 

The pairwise Margin-MSE training \cite{hofstaetter2020_crossarchitecture_kd} showed how pairwise knowledge distillation benefits from an ensemble of $\bertcat$ teachers. With tightly coupled teachers (TCT), \citet{lin2020distilling} showed the benefit of utilizing a $\colbert$ teacher model for in-batch negative signals. Both approaches add teacher inference overhead to the training. However, this can be mitigated by computing the teacher output once and re-using it.

\citet{ding2020rocketqa} showed with RocketQA a multi-generational process of training $\bertdot$ student and $\bertcat$ filtered negative passage sampling. They also showed how a very large batch size of 4,000 leads to large gains in accuracy on the sparse MSMARCO-DEV labels. Combined they require an enormous compute capacity (as the batch size has to fit into the GPU memory simultaneously) and time requirement for training a single instance.

%
%
Apart from specifically training dense retrieval models, knowledge distillation has gained popularity, with general-purpose BERT-style models \cite{jiao2019tinybert,sanh2019distilbert} as well as a range of applications in IR: from sequential recommendation models \cite{tang2018ranking}, BERT-based retrieval chatbots \cite{vakili2020distilling}, BERT-based Question Answering \cite{izacard2020distilling}, reducing the size of the $\bertcat$ passage re-ranking model \cite{gao2020understanding,chen2020simplified}, to dense keyword matching in sponsored search \cite{lu2020twinbert}.

%
%
The composition or sampling of training batches spans all machine learning application fields. Many advances were made especially in computer vision: whether to create synthetic negative samples for contrastive learning \cite{kalantidis2020hard}, 
unsupervised image cluster learning \cite{caron2021unsupervised}, or 
changing the image mixture of batches for self-supervised representation learning \cite{shen2020rethinking}. In IR, \citet{cohen2019} demonstrated that the sampling policy for negative samples plays an important role in the stability of the training, and \citet{macavaney2020training} adapted the training procedure by shifting samples to the beginning which are estimated to be easy.

\begin{figure*}[t]
   \includegraphics[trim={0.3cm 0.3cm 0.3cm 0.7cm},clip,width=0.9\textwidth]{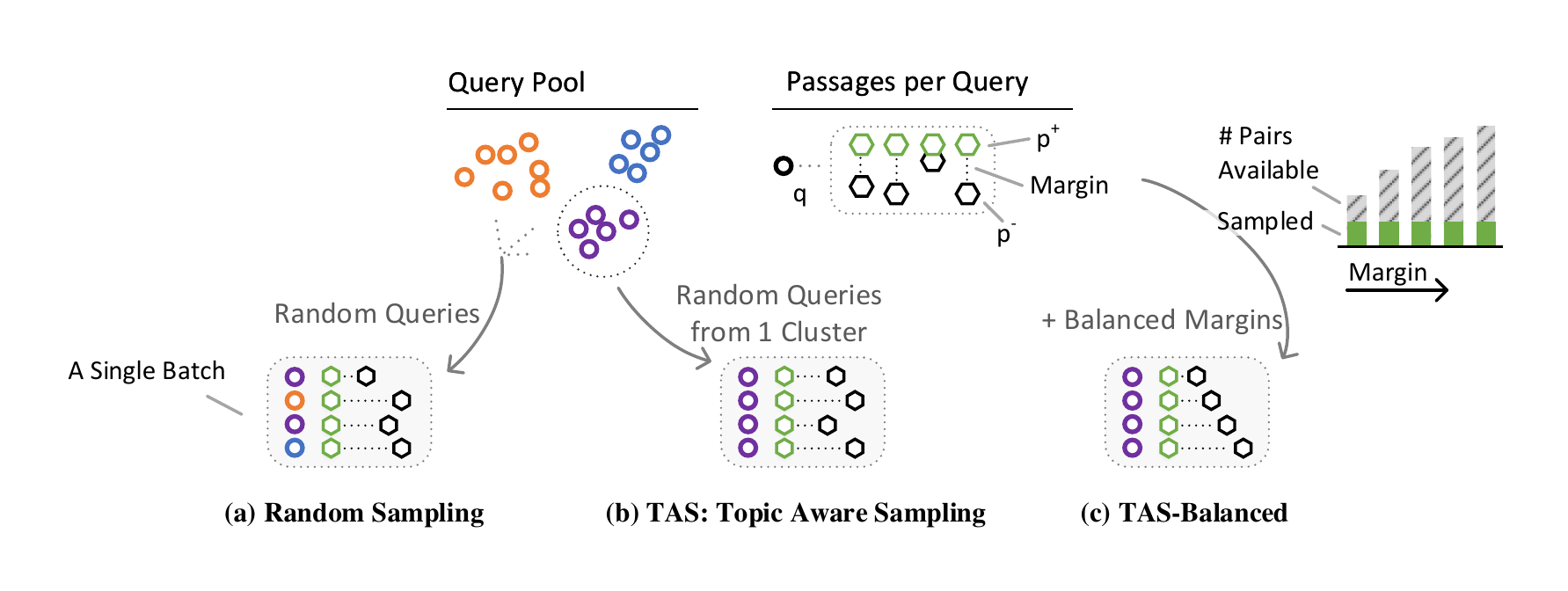}
    \centering
    \vspace{-0.6cm}
    \caption{Comparison of batch sampling strategies. Each strategy has access to a pool of (clustered) queries; where each query has a set of relevant and non-relevant passage pairs with BERT$_\text{CAT}$ score margins.}
    \label{fig:sampling_strategy}
   \vspace{-0.3cm}
\end{figure*}

\section{Topic Aware Sampling}
\label{sec:topic-sampling}

Clustering data has a long history in IR, for example in the cluster hypothesis concerned with retrieving clustered documents \cite{jardine1971use,voorhees1985cluster}. Inspired by these fundamental findings, we turn to clustering queries, as it is much more efficient than clustering passages, because we have fewer queries than passages available in the MSMARCO training data and each query is more rapidly encoded. We cluster queries to sample out of clusters for our topic aware training batches. We balance the passage pair selection to cover close and distant passage pairs uniformly. This reduces the amount of high-margin (low information) passage pairs. We combine the sampling with an efficient dual-teacher supervision method that combines pairwise and in-batch negative teachers. 

Typically, neural networks trained using gradient descent methods consider  a collection of training samples together as a batch, for a single update to the network parameters. The commonly used approach to compose such a batch $\mathcal{B}$ with size $b$ in the retrieval task is to take a sample of a query $q$, a relevant $p^+$ and a non-relevant $p^-$ passage randomly from the training data of all queries $Q$ and passage pairs per query $P_q$, as shown in Figure \ref{fig:sampling_strategy} (a). Formally:
\begin{equation}
\begin{aligned} 
\mathcal{B} = \big\{(q,p^+,p^-) \ \big\vert \ &q \in \textit{rand}(Q,b), \\&p^+,p^- \in \textit{rand}(P_q) \big\}
\end{aligned}
\end{equation}
where $\textit{rand}(X,y)$ is a random sampling method of $y$ samples (1 if omitted) from the population $X$ without replacement.
With hundreds of thousands of possible training queries, this random sampling produces a collection of queries per batch that cover completely different topics. Using an in-batch negative loss, where each query interacts not only with its own passages, but all others in the batch as well, has very little information gain from those random in-batch interactions. In-batch negatives offer a great promise of ``re-using'' already computed representations in the loss function.  

\smallskip \noindent {\bf TAS.} To fulfill the promise of improved training with in-batch negatives, we propose a Topic Aware Sampling (TAS) strategy, as depicted in Figure \ref{fig:sampling_strategy} (b). Before training, we group all training queries into $k$ clusters with k-means clustering \cite{macqueen1967some}, using their baseline representation vectors and minimize:
\begin{equation}
\begin{aligned} 
\argmin_C \  \sum_{i=1}^{k} \sum_{q \in C_i} || q - v_i ||^2
\end{aligned}
\end{equation}
where $v_i$ is the centroid vector of the group $C_i$. The results of this one-time and very efficient procedure are topically related clusters, as shown in the example Figure \ref{fig:topic_hero}. Now, instead of randomly selecting queries out of the full pool of queries, we randomly select $\lfloor b/n \rfloor$ queries out of $n$ random clusters from $C$ to create a batch:
\begin{equation}
\begin{aligned} 
\mathcal{B} = \big\{(q,p^+,p^-) \ \big\vert \ &
\begingroup
      \color{niceBlue}\overbrace{\color{black}{q \in \textit{rand}(\textit{rand}(C,n),\lfloor b/n \rfloor)}}^{\mathclap{\text{Topic Aware Sampling}}}
\endgroup
\\&p^+,p^- \in \textit{rand}(P_q) \big\}
\end{aligned}
\end{equation}

\smallskip \noindent {\bf TAS-Balanced.} As a further refinement, we augment TAS with the need to balance the pairwise margins. Naturally, most queries have fewer relevant passages than non-relevant ones \cite{zobel1998reliable}. We define negative passages to be \textit{easy} when they are further away from the positive passage in terms of the teacher model margin. To create a balanced sampling, based on the static margins of the pairwise teacher model $M_t$, as shown in Figure \ref{fig:sampling_strategy} (c), we define a method $H$ that filters passage pairs based on $h$ ranges of size $m$, that uniformly cover the minimum $m_{min}$ to the maximum margin per query:
\begin{equation}
\begin{aligned} 
H(P_{q},i) = \big\{(p^+,p^-) \ \big\vert \ & m_{min} + i \times m \\ &\geq M_t(q,p^+) - M_t(q,p^-) < \\& m_{min} + (i+1) \times m \big\}
\end{aligned}
\end{equation}
Similar to sampling clusters first, and then queries, we sample a range first and then sample from the filtered pairs to unskew the distribution of sample passage pairs.
Together, this yields our TAS-Balanced batch sampling strategy:
\begin{equation}
\begin{aligned} 
\mathcal{B} = \big\{(q,p^+,p^-) \ \big\vert \ &
\begingroup
      \color{niceBlue}\overbrace{\color{black}{q \in \textit{rand}(\textit{rand}(C,n),b)}}^{\mathclap{\text{Topic Aware Sampling}}}
\endgroup
, \\&\begingroup
      \color{niceRed}\underbrace{\color{black}{p^+, p^- \in \textit{rand}(H(P_{q},\textit{rand}(0..h)))}}_{\mathclap{\text{Balanced Margin Sampling}}}
\endgroup
\big\}
\end{aligned}
\end{equation}
The random sampling does not slow down our training loop as we conduct this batch composition concurrently in a sub-process and queue batches. For training we continuously sample new batches and do not repeat the same batch in multiple epochs. Rather than training for a certain number of epochs, our early-stopping approach detailed in Section \ref{sec:earlystop} decides when to stop training.

\vspace{-0.2cm}
\section{Experiment design}

Our main training and inference dependencies are PyTorch~\cite{pytorch2017}, HuggingFace Transformers \cite{wolf2019huggingface}, and Faiss \cite{faiss2017}, which we use for query clustering as well as brute-force nearest neighbor retrieval. 
\vspace{-0.2cm}
\subsection{Passage Collection \& Query Sets}

We use the MSMARCO-Passage~\cite{msmarco16} collection with the sparsely-judged MSMARCO-DEV query set of 6,980 queries (used in the leaderboard) as well as the densely-judged query sets of 43 and 54 queries derived from TREC-DL '19 \cite{trec2019overview} and '20 \cite{trec2020overview}. For TREC graded relevance (0 = non relevant to 3 = perfect) we use the recommended binarization point of 2 for MRR, MAP, and recall. MSMARCO is based on sampled Bing queries and contains 8.8 million passages. We use the official BM25-based 40 million training passage-pair samples. We cap the query length at $30$ tokens and the passage length at $200$ tokens; both values represent generous bounds with few outliers that have more tokens.

\subsection{Parameter Settings}

For our TAS clustering we use a pairwise trained $\bertdot$ model baseline as our source for the query representations. We create 2K clusters from the 400K training queries. A pilot study did not find any noticeable difference in the number of clusters. We set the number $n$ of clusters to sample from to 1 due to our relatively low batch size $b$ of 32 (if not further specified). We balance the margin ranges into 10 bins ($h$). After a pilot study we set the dual-teacher combination hyperparameter $\alpha$ to $0.75$ to bring both losses into the same range, as the in-batch loss, taking into account more data points, is consistently higher than the pairwise loss. We use the Adam \cite{kingma2014adam} optimizer with a learning rate of $7 \times 10^{-6}$.

As a basis for all our $\bertdot$ and $\colbert$ instances we use a 6-layer DistilBERT \cite{sanh2019distilbert} encoder as their initialization starting point. Each instance starts fresh from this DistilBERT checkpoint; we do not use generational retrieval-trained model checkpoints. We trained our $\colbert$ teacher model with the teacher pairwise signals. While we always used the static pairwise signals, due to studying many different batch compositions we implemented the $\colbert$ in-batch teacher as a dynamic sub-process running either on the same GPU for a batch size of 32 or an independent GPU for batch size 96 and 256. For the BM25 baseline we use Anserini \cite{Yang2017}.

\subsection{Approximate Retrieval Early Stopping}
\label{sec:earlystop}

We aim to train our neural retrieval models for as long as training improves the model, to facilitate the fairest comparison of our baselines and novel contributions. This is to not disadvantage methods that might take longer to train, but eventually catch up. 
We created an approximated early stopping set for all our experiments by indexing a pairwise trained baseline model and retrieving the top 100 passages for 3,200 queries uniformly sampled from the larger DEV-49K set, which are distinct from the DEV-7K and TREC evaluation sets. Additionally, we added all relevant passages if they have not been retrieved by the baseline already. 
Evaluating our early stopping set takes 5 minutes and we evaluate it every 4K steps; we stop training a model after 30 evaluations have not improved the nDCG@10 metric, which usually stops after 700-800K steps.

\section{Results}

In this section we discuss our research questions and present our results: We compare to internal baselines of different teacher and sampling modes; compare our results to external baselines; study the robustness of our TAS method; and finally provide insights into the use of TAS in a broader re-ranking system. Except for the last point, we present results of the trained $\bertdot$ model using a nearest neighbor search across all MSMARCO passages without re-ranking or fusion of results. 

\subsection{Source of Effectiveness}
\label{sec:ablations}
With our proposal to change the batch sampling on one hand and the teacher supervision on the other, we carefully study the effects of each change in:
\RQone

\begin{table}[t]
    \centering
    \caption{Analysis of TAS-Balanced \& dual-supervision using different loss methods for in-batch negative signals. nDCG \& MRR cutoff 10.  \textit{Stat.\ sig.\ difference w/ paired t-test (p < 0.05)}}
    \label{tab:loss_abl}
    \setlength\tabcolsep{2.7pt}
    \vspace{-0.3cm}
    \begin{tabular}{l!{\color{lightgray}\vrule}ll!{\color{lightgray}\vrule}ll!{\color{lightgray}\vrule}ll}
       
       \toprule

       \multirow{2}{*}{\textbf{Loss}} &
       \multicolumn{2}{c!{\color{lightgray}\vrule}}{\textbf{TREC-DL'19}}&
       \multicolumn{2}{c!{\color{lightgray}\vrule}}{\textbf{TREC-DL'20}}&
       \multicolumn{2}{c}{\textbf{MSM. DEV}}\\
       & {\small nDCG} & {\small R@1K} & {\small nDCG} & {\small R@1K} & {\small MRR} & {\small R@1K} \\

       \midrule
       \arrayrulecolor{lightgray}
        \textbf{\underline{K}LDiv}       & .681  & .783  & .673  & .831  & .334  & .964  \\ 
        \textbf{List\underline{N}et}     & .687  & .788  & .668  & .829  & .338$^{k}$  & .966$^{k}$  \\ 
        \textbf{\underline{L}ambdarank}  & .704  & .812$^{kn}$  & .682  & .840  & \textbf{.342}$^{kn}$  & .971$^{kn}$  \\ 
        \textbf{\underline{M}argin-MSE}  & \textbf{.712}  & \textbf{.845}$^{kln}$  & \textbf{.693}  & \textbf{.865}$^{kln}$  & .340$^{k}$  & \textbf{.975}$^{kln}$  \\ 
        \arrayrulecolor{black}
       \bottomrule
    \end{tabular}
        \vspace{-.5cm}
\end{table}
\begin{table*}[t]
    \centering
    \caption{Ablation results of random, TAS, and TAS-Balanced sampling strategies. \textit{(paired t-test; p < 0.05)}}
    \label{tab:tas_teacher_abl}
    \setlength\tabcolsep{3pt}
    \vspace{-0.3cm}
    \begin{tabular}{ll!{\color{lightgray}\vrule}lll!{\color{lightgray}\vrule}lll!{\color{lightgray}\vrule}lll}
       
       \toprule

       \multirow{2}{*}{\textbf{Teacher}} &
       \multirow{2}{*}{\textbf{Sampling}}   &
       \multicolumn{3}{c!{\color{lightgray}\vrule}}{\textbf{TREC-DL'19}}&
       \multicolumn{3}{c!{\color{lightgray}\vrule}}{\textbf{TREC-DL'20}}&
       \multicolumn{3}{c}{\textbf{MSMARCO DEV}}\\
       && nDCG@10 & MRR@10 & R@1K & nDCG@10 & MRR@10 & R@1K & nDCG@10 & MRR@10 & R@1K \\

       \midrule
       \arrayrulecolor{lightgray}
        None & Random & .602 & .781 & .714 & .602 & .782 & .757 & .353 & .298 & .935 \\
       \midrule
       \multirow{3}{*}{Pairwise ($\bertcat$)} & \underline{R}andom        & .687 & .851 & .767 & .654 & .812 & .801 & .385 & .326 & .958 \\
                                              & \underline{T}AS           & .677 & .851 & .769 & .650 & .820 & .819 & .385 & .325 & .957 \\
                                              & TAS-\underline{B}alanced  & .686 & .866 & .783 & .665 & .823 & .825$^{r}$ & .393$^{rt}$ & .334$^{rt}$ & .963$^{r}$ \\
       \midrule
       \multirow{3}{*}{In-Batch Neg. ($\colbert$)} & \underline{R}andom        & .680 & .857 & .745 & .631 & .773 & .792 & .372 & .315 & .951 \\
                                                   & \underline{T}AS           & .706 & .886 & .799 & .667$^{r}$ & .821 & .826$^{r}$ & .396$^{r}$ & .336$^{r}$ & .968$^{r}$ \\
                                                   & TAS-\underline{B}alanced  & \textbf{.716} & \textbf{.910} & .800 & .677$^{r}$ & .810 & .820$^{r}$ & .397$^{r}$ & .338$^{r}$ & .968$^{r}$ \\

       \midrule
       \multirow{3}{*}{Pairwise + In-Batch} & \underline{R}andom         & .695 & .891 & .787 & .673 & .812 & .839 & .391 & .331 & .968 \\
                                            & \underline{T}AS            & .713 & .878 & .831 & .689 & .815 & .862$^{r}$ & .401$^{r}$ & .338$^{r}$ & .973$^{r}$ \\
                                            & TAS-\underline{B}alanced   & .712 & .892 & \textbf{.845} & \textbf{.693} & \textbf{.843} & \textbf{.865}$^{r}$ & \textbf{.402}$^{r}$ & \textbf{.340}$^{r}$ & \textbf{.975}$^{rt}$ \\
        \arrayrulecolor{black}
       \bottomrule
    \vspace{-.7cm}
    \end{tabular}
\end{table*}

\noindent For pairwise knowledge distillation, \citet{hofstaetter2020_crossarchitecture_kd} showed their proposed Margin-MSE loss to outperform other options, therefore we fix the use of the Margin-MSE loss for the pairwise teaching part and examine the effect of different losses for additional in-batch negative training for our TAS-Balanced strategy in Table \ref{tab:loss_abl}. 

We study two different types of loss functions: First are list-based losses  (KL Divergence, ListNet \cite{cao2007learning}, and the LambdaLoss version nDCG2 \cite{wang2018lambdaloss}) where we build a ranked list of in-batch and pairwise negatives per query and second the pairwise Margin-MSE loss that repeats the relevant passage per query to pair with all other passages from the in-batch negative pool of a batch. 

We find in Table \ref{tab:loss_abl} that the Margin-MSE loss outperforms other list-based loss variants in most metrics across our three query sets. The change is especially noticeable in the recall metric on the two TREC-DL query sets. The Margin-MSE loss in comparison to the list-based losses optimizes the $\bertdot$ model to follow the teacher score distribution and not just the general ordering of in-batch negatives. We hypothesize the reason for the better Margin-MSE results is because it is advantageous to use a homogeneous loss between both teachers and, because list-based losses only observe ordering and the in-batch negatives are still an incomplete set of all available orderings, the score optimization is more precise.

Our main ablation results in Table \ref{tab:tas_teacher_abl} investigate two axes: the type of teacher supervision and the type of sampling with all possible combinations between our proposed methods. We also provide a baseline of a standalone-trained $\bertdot$ model with random batch sampling and binary relevance labels only. 

The first teacher scenario uses only the pairwise teacher ensemble scores. Comparing the pairwise teacher with the standalone model, we already see significant gains over all metrics. TAS sampling alone does not change the results much and even decreases TREC results slightly. This is an expected result, as the TAS sampling is geared towards in-batch negative training, and should not strongly influence training on queries independently. The TAS-Balanced procedure, on the other hand, improves most metrics for pairwise teaching by 1 percentage point or more, as the balanced margin sampling influences the pairwise supervision. 

Using in-batch negatives and a single $\colbert$ teacher model for supervision with the Margin-MSE loss shows worse results for the original random sampling than the pairwise teacher on the same setting. Here, the TAS strategy provides a strong boost for the results, across all three collections. The TAS-Balanced strategy again improves results for two of the three collections. 

Finally, using our novel dual-supervision strategy we observe the same pattern again: TAS improves over random sampling, and TAS-Balance improves over TAS for the best results on almost any evaluated metric. When we look at the differences between the in-batch teaching and the dual-teacher we see that, especially on the recall metrics, the dual-teacher outperforms the single teacher by a large margin on all three query sets. The nDCG and MRR results are improved for dual-supervision on two out of the three query sets and the remaining TREC-DL'19 results are tied. 
Because of these results, we recommend using the dual-supervision and TAS-Balanced sampling as the main configuration and we use it throughout the paper for our analysis.

TAS-Balanced uses randomized sampling out of cluster, query, and passage-pair populations extensively. To be confident in our results, we need to investigate if we did inadvertently overfit our approach to a certain setting and study: 

\RQtwo

\noindent In Table \ref{tab:rand_abl} we present the results of our robustness analysis for TAS-Balanced and dual-supervision with different random seeds that guide the ordering and selection of the training samples. Every instance had access to the same data, training configuration, teacher models -- the only difference is the random ordering of clusters, queries and passage-pairs. We find overall low variability in our results, especially on the 6,980 test queries of MSMARCO-DEV. For the TREC-DL sets we have many fewer queries -- 43 and 53 for TREC-DL'19 and '20 respectively -- and still our robustness analysis shows a standard deviation of the results under a single point change in both nDCG@10 and recall. The biggest variation is on the nDCG@10 metric of TREC-DL'20, however the recall shows a lower variance than the recall on TREC-DL'19. This result gives us great confidence in the efficacy of our TAS-Balanced training.

\begin{table}[t]
    \centering
    \caption{Random-robustness analysis of five instances of TAS-Balanced dual-supervision each using different sampling orders across clusters, queries, and passage pairs. \textit{Stat.\ sig.\ difference w/ paired t-test (p < 0.05)}}
    \label{tab:rand_abl}
    \setlength\tabcolsep{1.7pt}
    \vspace{-0.3cm}
    \begin{tabular}{l!{\color{lightgray}\vrule}ll!{\color{lightgray}\vrule}ll!{\color{lightgray}\vrule}ll}
       
       \toprule

       \multirow{2}{*}{\textbf{Inst.}} &
       \multicolumn{2}{c!{\color{lightgray}\vrule}}{\textbf{TREC-DL'19}}&
       \multicolumn{2}{c!{\color{lightgray}\vrule}}{\textbf{TREC-DL'20}}&
       \multicolumn{2}{c}{\textbf{MSMARCO DEV}}\\
       & nDCG@10 & R@1K & nDCG@10 & R@1K & MRR@10 & R@1K \\

       \midrule
       \arrayrulecolor{lightgray}
        \textbf{A}  & .712  & .845  & .693  & .865$^{bc}$  & .340  & .975  \\ 
        \textbf{B}  & .713  & .833  & .684  & .859  & .341  & .974  \\ 
        \textbf{C}  & .716  & .844  & .679  & .859  & .341  & .975$^{b}$  \\ 
        \textbf{D}  & .712  & .838  & .688  & .861  & .339  & .974  \\ 
        \textbf{E}  & .705  & .841  & .701$^{bc}$  & .862  & .339  & .974  \\ 
        \midrule
        \textbf{Avg.}      & .712  & .840  & .689  & .861   & .340  & .975  \\
        \textbf{StdDev.}  & .004  & .005  & .008  & .003   & .001  & .001  \\
        \arrayrulecolor{black}
       \bottomrule
    \end{tabular}
    \vspace{-.5cm}
    
\end{table}

\begin{table*}[t!]
    \centering
    \caption{Dense retrieval results of $\bertdot$ for baseline training and our TAS-Balanced training. \textit{L\# = Transformer layers} \textit{Stat.\ sig.\ difference w/ paired t-test (p < 0.05) b=BM25; T=TCT; M=Margin-MSE; 3=TAS-B 32; 9=TAS-B 96; 2=TAS-B 256}}
    \label{tab:dr_results}
    \vspace{-0.4cm}
    \setlength\tabcolsep{.8pt}
    \begin{tabular}{lllll!{\color{lightgray}\vrule}lll!{\color{lightgray}\vrule}lll!{\color{lightgray}\vrule}lll}
       \toprule
       &\multirow{2}{*}{\textbf{Training Type}}   &
       \multirow{2}{*}{\textbf{Encoder}} & \multirow{2}{*}{\textbf{L\#}} &
       \textbf{Batch} &
       \multicolumn{3}{c!{\color{lightgray}\vrule}}{\textbf{TREC-DL'19}}&
       \multicolumn{3}{c!{\color{lightgray}\vrule}}{\textbf{TREC-DL'20}}&
       \multicolumn{3}{c}{\textbf{MSMARCO DEV}}\\
       &&&&\textbf{Size}& nDCG@10 & MRR@10 & R@1K & nDCG@10 & MRR@10 & R@1K & nDCG@10 & MRR@10 & R@1K \\
        \midrule
        \multicolumn{6}{l}{\textbf{Baselines}} \\
         &\underline{B}M25 & -- & --    & --                                                                     & .501 & .689 & .739 & .475 & .649 & .806 & .241 & .194 & .868 \\
         
         \arrayrulecolor{lightgray}
         \midrule
        \cite{xiong2020approximate} &ANCE  & \multirow{3}{*}{BERT-Base} & \multirow{3}{*}{12}                                 & \multirow{3}{*}{32}    & .648 & -- & -- & -- & -- & -- & -- & .330 & .959 \\
         \cite{zhan2020learning}&LTRe  &  &                                      &     & .661 & -- & -- & -- & -- & -- & -- & .329 & .955 \\
         \cite{zhan2020learning}&ANCE + LTRe  &  &   &                                 & .675 & -- & -- & -- & -- & -- & -- & .341 & .962 \\
         \midrule
         \multirow{2}{*}{\cite{ding2020rocketqa}} &\multirow{2}{*}{RocketQA}  & \multirow{2}{*}{ERNIE-Base} & \multirow{2}{*}{12}                       & 4,000 &   -- & -- & -- & -- & -- & -- & -- & \textbf{.364} & -- \\
         &&  &                        & 128   &   -- & -- & -- & -- & -- & -- & -- & .309 & -- \\ 
         \midrule
         \multirow{2}{*}{\cite{lin2020distilling}} & TCT  & BERT-Base & 12                                & 96    & .670 & -- & .720 & -- & -- & -- & -- & .335 & .964 \\
         &\underline{T}CT (ours) & DistilBERT & 6                                                    & 32    & .680$^{b}$ & .857$^{b}$ & .745 & .631$^{b}$ & .773$^{b}$ & .792 & .372$^{b}$ & .315$^{b}$ & .951$^{b}$ \\
         \midrule
         \multirow{2}{*}{\cite{hofstaetter2020_crossarchitecture_kd}}&Margin-MSE  & \multirow{2}{*}{DistilBERT} & \multirow{2}{*}{6}        & \multirow{2}{*}{32}     & .697 & .868 & .769 & -- & -- & -- & .381 & .323 & .957 \\
         &\underline{M}argin-MSE (ours) &  &                                              &      & .687$^{b}$ & .851$^{b}$ & .767 & .654$^{b}$ & .812$^{b}$ & .801 & .385$^{bt}$ & .326$^{bt}$ & .958$^{bt}$ \\
        \arrayrulecolor{black}
        \midrule
         \multicolumn{6}{l}{\textbf{Ours}} \\
         \arrayrulecolor{lightgray}
         
         &\multirow{3}{*}{TAS-Balanced} & \multirow{3}{*}{DistilBERT } & \multirow{3}{*}{6}  & \underline{3}2     & .712$^{b}$ & .892$^{b}$ & \textbf{.845}$^{btm}$ & \textbf{.693}$^{btm}$ & \textbf{.843}$^{b}$ & .865$^{btm}$ & .402$^{btm}$ & .340$^{btm}$ & .975$^{btm}$  \\
                                                                                        & & && \underline{9}6     & \textbf{.722}$^{btm}$ & \textbf{.895}$^{b}$ & .842$^{tm}$ & .692$^{btm}$ & .841$^{b}$ & .864$^{btm}$ & .406$^{btm}$ & .343$^{btm}$ & .976$^{btm}$  \\
                                                                                        && & &\underline{2}56     & .717$^{btm}$ & .883$^{b}$ & .843$^{tm}$ & .686$^{btm}$ & .843$^{b}$ & \textbf{.875}$^{btm}$ & \textbf{.410}$^{btm39}$ & .347$^{btm39}$ & \textbf{.978}$^{btm3}$ \\

        \arrayrulecolor{black}
        \bottomrule
    \end{tabular}
    \vspace{-0.3cm}
\end{table*}

\subsection{Comparing to Baselines}

In this section we focus on standalone $\bertdot$ retrieval results from different training methods and compare our results with related work to answer:

\RQthree

\noindent We present the dense retrieval results for models trained on the MSMARCO collection in Table \ref{tab:dr_results}, first the baselines and then our TAS-Balanced results using different training batch size settings. Important for the comparison of different $\bertdot$ training techniques is the number of Transformer encoder layers, which linearly determines the indexing throughput and query encoding latency, as well as the training batch size which influences the GPU memory requirements. The TREC-DL'20 query set was recently released, therefore most related work is missing results on these queries. We observe that the methods not using knowledge distillation and larger encoders (ANCE, LTRe) are outperformed on TREC-DL'19 by those that do use teachers (TCT, Margin-MSE), however on the sparse MSMARCO-DEV the result trend turns around. RocketQA on MSMARCO-DEV only outperforms all other approaches when using a batch size of 4,000; RocketQA using 128 samples per batch -- more than any other method, but the lowest published by the authors -- is outperformed by all other methods. 

Our TAS-Balanced results are in the last section of Table \ref{tab:dr_results}. We evaluated our 6 layer encoder model on three different training batch sizes (32, 96, and 256). Between the different batch sizes, we only see a clear trend of improvement on MSMARCO DEV, but not on the TREC collections, there the results are inconsistent, albeit with small differences, that fall in the standard deviation of our robustness analysis in Table \ref{tab:rand_abl}. This leads us to believe that increasing the batch size is a source of overfitting on the sparse MSMARCO labels. 
Our TAS-Balanced models outperform all other dense retrieval training methods on both TREC-DL query sets, which show very similar trends: nDCG@10 by at least 4\%; MRR@10 by 3\%; and Recall@1K, where the margin is the highest with at least 9\% improvement over the respectively best related work baseline. TAS-Balanced also shows consistently strong results on the sparse MSMARCO DEV, where we outperform all other baselines, especially on Recall@1K. The only stronger baseline is RocketQA's 4,000 batch size instance; however, as we discussed, this is only due to the larger batch size and not because of their approach, as we strongly outperform (+10\% on MRR@10) their 128 batch size instance with a batch size as low as 32.

At this point we want to take a step back and examine the results from a perspective before the \textit{neural revolution}: Our TAS-Balanced trained $\bertdot$ dense retriever, which has comparable query latency with BM25, outperforms BM25 by 44\% on nDCG@10 and 9-14\% on Recall@1K on TREC'19 \& '20. Our work is only the latest in an enormous progress the community made the last few years.

\subsection{TAS-Balanced Retrieval in a Pipeline}

Although the quality of our top-10 results allows use of our $\bertdot$ model as a standalone retriever, usually a ranking system is a hybrid combining different relevance signals. Thus, we investigate:

\RQfour

\noindent We create two pipelines: First, we fuse our TAS-Balanced retriever with docT5query, a sparse passage expansion based retriever, following the setup of \citet{lin2020distilling}. Second, we re-rank our results with the state-of-the-art mono-duo-T5 re-ranking model following \citet{pradeep2021expando}.

\begin{figure}[t]
    \centering
    \includegraphics[width=0.36\textwidth,clip, trim=0.2cm 0.2cm 0.2cm 0.2cm]{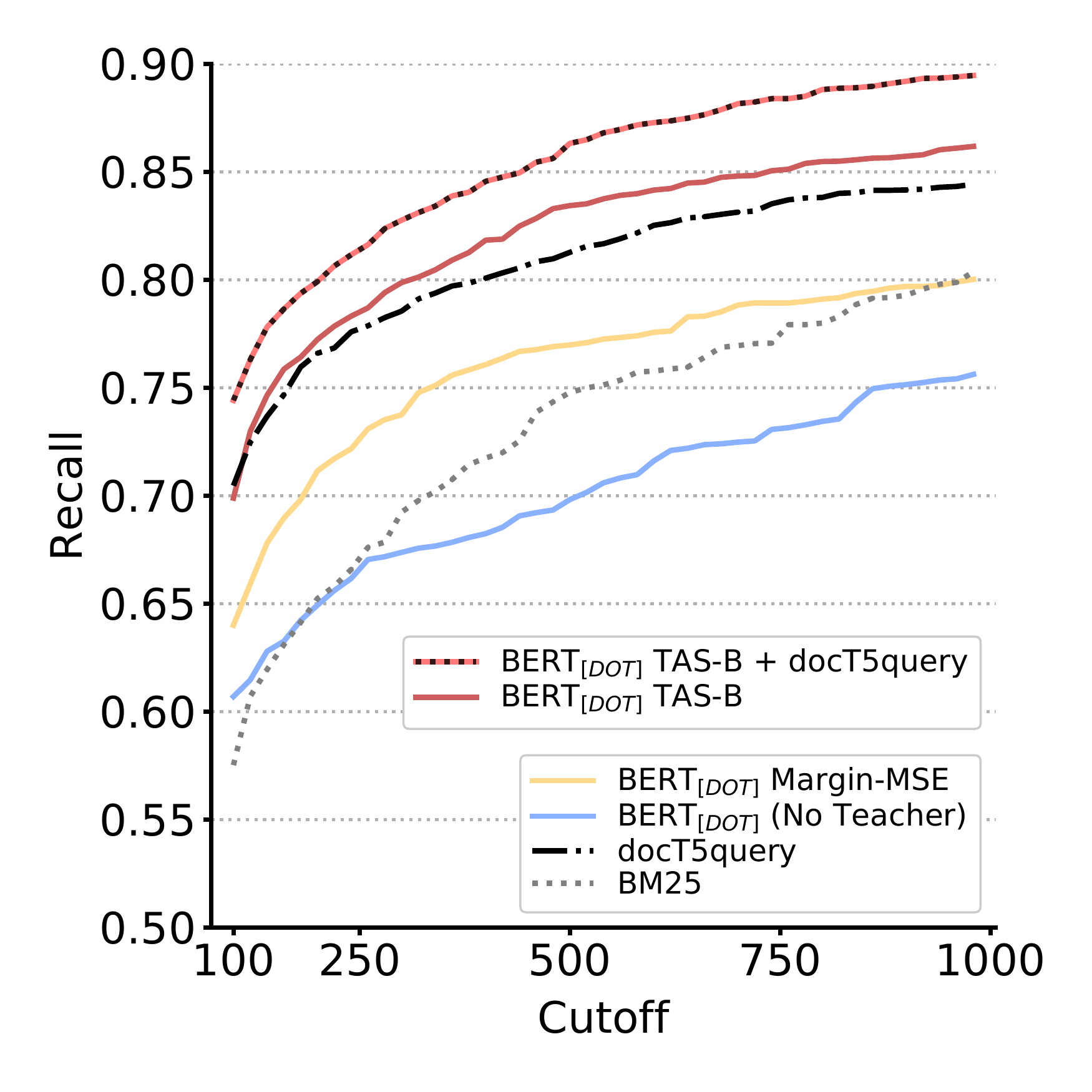}
    \vspace{-0.6cm}
  \caption{Recall at different cutoffs for TREC-DL'20}
  \label{fig:recall-trec20-bin2}
  \vspace{-0.5cm}
\end{figure}

As a first step, we examine the usability of different first-stage retrievers in terms of their recall at different cutoffs in Figure \ref{fig:recall-trec20-bin2}. These candidates can then be further used by re-ranking models. We find that on TREC-DL our dense retriever is the first dense retriever to consistently outperform BM25 and docT5query at all examined  cutoffs. The fused TAS-Balanced + docT5query results offer another boost of recall, showing us that those two diametrical methods bring different strengths that fit together very well. 

\begin{table*}[t!]
    \centering
    \caption{Full system results using retrieval and re-ranking pipelines. \textit{Stat.\ sig.\ difference w/ paired t-test (p < 0.05)}}
    \label{tab:hybrid_pipe_results}
    \vspace{-0.4cm}
    \setlength\tabcolsep{1.5pt}
    \begin{tabular}{lllrr!{\color{lightgray}\vrule}lll!{\color{lightgray}\vrule}lll!{\color{lightgray}\vrule}lll}
       \toprule
       &\multirow{2}{*}{\textbf{Retrieval-Stage}}   &
       \multicolumn{2}{c}{\textbf{Re-ranking}} &
       \textbf{Latency} &
       \multicolumn{3}{c!{\color{lightgray}\vrule}}{\textbf{TREC-DL'19}}&
       \multicolumn{3}{c!{\color{lightgray}\vrule}}{\textbf{TREC-DL'20}}&
       \multicolumn{3}{c}{\textbf{MSMARCO DEV}}\\
       &&{Model}&\#&(ms)& {\small nDCG@10} & {\small MRR@10} & {\small R@1K} & {\small nDCG@10} & {\small MRR@10} & {\small R@1K} & {\small nDCG@10} & {\small MRR@10} & {\small R@1K} \\
        \midrule
        \multicolumn{7}{l}{\textbf{Low Latency Systems (<70ms)}} \\
        \cite{Yang2017} & \underline{B}M25 & -- & --    & 55                                        & .501 & .689 & .745 & .475 & .649 & .803 & .241 & .194 & .857 \\
        \cite{dai2019context} &DeepCT & -- & --    & 55                                 & .551 & -- & .756 & -- & -- & -- & -- & .243 & .913 \\
        \cite{nogueira2019doct5query} &\underline{d}ocT5query & -- & --    & 64                     & .648$^{b}$ & .799 & .827 & .619$^{b}$ & .742 & .844$^{b}$ & .338$^{b}$ & .277$^{b}$ & .947$^{b}$ \\
        
        \arrayrulecolor{lightgray}
        \midrule
        
          &\underline{T}AS-B                                                     & -- & -- &   64   & .722$^{bd}$ & .895$^{b}$ & .842 & .692$^{bd}$ & \textbf{.841}$^{bd}$ & .864$^{b}$ & .406$^{bd}$ & .343$^{bd}$ & .976$^{bd}$  \\
          &T\underline{A}S-B  + docT5query                                       & -- & -- &   67   & \textbf{.753}$^{bdt}$ & \textbf{.920}$^{bd}$ & \textbf{.882}$^{bdt}$ & \textbf{.708}$^{bd}$ & .832$^{b}$ & \textbf{.895}$^{bdt}$ & \textbf{.425}$^{bdt}$ & \textbf{.360}$^{bdt}$ & \textbf{.979}$^{bdt}$  \\

        \arrayrulecolor{black}
        \midrule
        \multicolumn{7}{l}{\textbf{Medium Latency Systems (< 500ms)}} \\
        \arrayrulecolor{lightgray}
        
        \cite{lin2020distilling} & TCT + docT5query  & -- & --                 & 106   & .739 & -- & .832 & -- & -- & -- & -- & .364 & .973 \\
        \cite{khattab2020colbert} & ColBERT & -- & --                          &  458   & -- & -- & -- & -- & -- & -- & -- & .360 & .968 \\
        \cite{zhan2020learning} & LTRe  & BERT-Large & 10                      &  148   & -- & -- & -- & -- & -- & -- & -- & .362 & .962 \\
          \multirow{1}{*}{\cite{pradeep2021expando}}&\multirow{1}{*}{\underline{B}M25} & \multirow{1}{*}{duo-T5} & 10       &  388   & .553 & .839 & .745 & .544 & .793 & .803 & .310 & .287 & .857 \\
         \multirow{1}{*}{\cite{pradeep2021expando}}&\multirow{1}{*}{\underline{d}ocT5query} & \multirow{1}{*}{duo-T5} & 10  &   397  & .696$^{b}$ & \textbf{.913} & .827 & .658$^{b}$ & .839 & .844$^{b}$ & .411$^{b}$ & .371$^{b}$ & .947$^{b}$ \\
         \midrule
         &\underline{T}AS-B & \multirow{1}{*}{duo-T5} & 10                                                                   &   397  & .727$^{b}$ & .877 & .842 & .710$^{b}$ & .864 & .864$^{b}$ & .449$^{bd}$ & .399$^{bd}$ & .976$^{bd}$ \\

         &T\underline{A}S-B + docT5query & \multirow{1}{*}{duo-T5}  & 10                                         &  400   & \textbf{.755}$^{bdt}$ & .877 & \textbf{.882}$^{bdt}$ & \textbf{.726}$^{bdt}$ & \textbf{.870} & \textbf{.895}$^{bdt}$ & \textbf{.463}$^{bdt}$ & \textbf{.409}$^{bdt}$ & \textbf{.979}$^{bdt}$ \\

        \arrayrulecolor{black}
        \midrule
         \multicolumn{7}{l}{\textbf{High Latency Systems (> 500ms)}} \\
         \arrayrulecolor{lightgray}
         \cite{nogueira2019passage} & BM25 & BERT-Large  & 1K                  &  3,500 & .736 & -- & -- & -- & -- & -- & -- & .365 & -- \\
         \cite{pradeep2021expando}  & \underline{B}M25&          mono-duo-T5   & 1K        &  12,800 & .760 & .852 & .745 &.774 & \textbf{.888} & .803 & .471 & .409 & .857 \\
         \cite{pradeep2021expando}  & \underline{d}ocT5query &  mono-duo-T5    & 1K         &  12,800 & \textbf{.773} & \textbf{.864} & .827 & \textbf{.784}$^{b}$ & .880 & .844$^{b}$ & .488$^{b}$ & .420 & .947$^{b}$ \\

         \midrule
                  &\underline{T}AS-B&   mono-duo-T5  & 1K                        &  12,800    & .759 & .846 & .842 & .782 & .881 & .864$^{b}$ & .488$^{b}$ & .420  & .976$^{bd}$  \\
                  &T\underline{A}S-B + docT5query&        mono-duo-T5  & 1K      &  12,800   & .759 & .848 & \textbf{.882}$^{bdt}$ & .783 & .880 & \textbf{.895}$^{bdt}$ & \textbf{.489}$^{bt}$ & \textbf{.421}$^{d}$ & \textbf{.979}$^{bdt}$  \\

        \arrayrulecolor{black}
        \bottomrule
    \end{tabular}
    \vspace{-0.4cm}
\end{table*}

In Table \ref{tab:hybrid_pipe_results} we compare our multi-stage pipelines grouped by latency.
Naturally, the higher the re-ranking depth the higher the full system latency. 
Analyzing the baselines, we see a large spread in terms of per query latency and effectiveness results. 
Low-latency results are generally inferior in terms of quality compared to the slower re-ranking results. 
The powerful re-ranking models are able to improve results when the re-ranking depth is as small as 10 candidates (especially on TREC'20 and MSMARCO-DEV), albeit they show a larger improvement for 1,000 candidates.

Turning to our results, we use the TAS-Balanced with dual-supervision trained on a batch size of 96 for all pipeline experiments.

%
%
\smallskip \noindent {\bf Low-Latency.} 
As with dense retrieval models (in Table \ref{tab:dr_results}), TAS-Balanced outperforms other low-latency (<70 ms) systems BM25, DeepCT, and docT5query by large margins in Table \ref{tab:hybrid_pipe_results}. 
Fusing TAS-Balanced together with docT5query further improves Recall@1K, as shown in Figure \ref{fig:recall-trec20-bin2}, as well as almost all results across query sets, at virtually no latency cost other than merging the two result lists.
Across every query set we show the highest Recall@1K with this fused first-stage, followed by our standalone TAS-Balanced retriever. 
The recall of these first-stage models naturally determines the recall for the re-ranking pipelines.
Comparing our low-latency TAS-Balanced (+ docT5query fusion) results with the medium-latency baselines, we observe that in many instances we already outperform or tie methods that are 2-6x slower.

%
%
\smallskip \noindent {\bf Medium-Latency.} 
As soon as we incorporate re-ranking models into a pipeline, we have an explosion of potential options, including the re-ranking depth. 
For medium-latency systems we re-rank only the top-10 candidates with the duo-T5 re-ranking model. 
While this top-10 approach only shows modest gains for TREC'19 on baselines and TAS-Balanced retrievers, the gains are much stronger on TREC'20 and MSMARCO-DEV. 
Following the low-latency pattern, our TAS-Balanced (+ docT5query fusion) re-ranked with duo-T5 outperform other duo-T5 re-ranking pipelines as well as other related systems such as ColBERT or a BERT-large re-ranking system.

%
%
\smallskip \noindent {\bf High-Latency.} 
Our final results employ the full mono-duo-T5 re-ranker at a depth of 1K, where mono-T5 re-ranks the 1K results and duo-T5 then scores the top-50.
This pipeline is hardly practical in a production scenario, with 13 seconds latency per query, but gives us a ceiling for the best achievable metrics with current state-of-the-art re-rankers. 
For MSMARCO-DEV our increased first-stage recall leads to slightly better re-ranking results than the first-stage baselines.
However, for the TREC query sets, even though TAS-Balanced shows a higher recall, the mono-duo-T5 re-ranker is (non significantly) better using BM25 \& docT5query as retriever. 
We believe that the mono-duo-T5 re-ranker is not able to take advantage of the increased recall because it has been trained on a BM25 candidate distribution and with dense retrieval we create a shift in the candidate distribution. 
It is out of the scope of this work to re-train the mono-duo-T5 re-rankers, albeit the importance of training the re-ranker on the first-stage retriever distribution is shown by \citet{gao2021rethink} and \citet{ding2020rocketqa}.
Overall, these are encouraging results to spark future pipeline work based on TAS-Balanced trained $\bertdot$ as a first-stage retrieval model.

\vspace{-0.3cm}
\section{Conclusion}

We proposed to improve dense passage retrieval training with a cost-neutral topic aware (query) and balanced margin (passage pairs) sampling strategy, called TAS-Balanced. We train the dual-encoder $\bertdot$ model with a dual-supervision of pairwise and in-batch teacher models. Our training only requires under 48 hours on a single consumer-grade GPU and outperforms most other approaches that depend on large server infrastructures, especially on two densely judged TREC-DL query sets. We showed TAS-Balanced works consistently with different random orderings and different teacher supervisions. Additionally, to our standalone retriever, we show how TAS-Balanced interacts with other models in a larger search pipeline. Using TAS-Balanced fused with docT5query results outperforms many systems with 2-6x higher latency. Furthermore, TAS-Balanced is also beneficial for low re-ranking depths. We purposefully set out to design a training technique for dense retrieval that does not depend on large compute servers. We want to give the community the techniques necessary to train strong dense retrieval models with modest hardware, so that the largest possible number of researchers and practitioners can benefit from the neural first-stage improvements and build upon them.

\balance
\bibliographystyle{ACM-Reference-Format}
\bibliography{references}


\begin{thebibliography}{45}


\ifx \showCODEN    \undefined \def \showCODEN     #1{\unskip}     \fi
\ifx \showDOI      \undefined \def \showDOI       #1{#1}\fi
\ifx \showISBNx    \undefined \def \showISBNx     #1{\unskip}     \fi
\ifx \showISBNxiii \undefined \def \showISBNxiii  #1{\unskip}     \fi
\ifx \showISSN     \undefined \def \showISSN      #1{\unskip}     \fi
\ifx \showLCCN     \undefined \def \showLCCN      #1{\unskip}     \fi
\ifx \shownote     \undefined \def \shownote      #1{#1}          \fi
\ifx \showarticletitle \undefined \def \showarticletitle #1{#1}   \fi
\ifx \showURL      \undefined \def \showURL       {\relax}        \fi
\providecommand\bibfield[2]{#2}
\providecommand\bibinfo[2]{#2}
\providecommand\natexlab[1]{#1}
\providecommand\showeprint[2][]{arXiv:#2}

\bibitem[\protect\citeauthoryear{Akkalyoncu~Yilmaz, Yang, Zhang, and
  Lin}{Akkalyoncu~Yilmaz et~al\mbox{.}}{2019}]%
        {yilmaz2019cross}
\bibfield{author}{\bibinfo{person}{Zeynep Akkalyoncu~Yilmaz},
  \bibinfo{person}{Wei Yang}, \bibinfo{person}{Haotian Zhang}, {and}
  \bibinfo{person}{Jimmy Lin}.} \bibinfo{year}{2019}\natexlab{}.
\newblock \showarticletitle{Cross-Domain Modeling of Sentence-Level Evidence
  for Document Retrieval}. In \bibinfo{booktitle}{\emph{Proc. of
  EMNLP-IJCNLP}}.
\newblock


\bibitem[\protect\citeauthoryear{Bajaj, Campos, Craswell, Deng, Gao, Liu,
  Majumder, McNamara, Mitra, and Nguyen}{Bajaj et~al\mbox{.}}{2016}]%
        {msmarco16}
\bibfield{author}{\bibinfo{person}{Payal Bajaj}, \bibinfo{person}{Daniel
  Campos}, \bibinfo{person}{Nick Craswell}, \bibinfo{person}{Li Deng},
  \bibinfo{person}{Jianfeng Gao}, \bibinfo{person}{Xiaodong Liu},
  \bibinfo{person}{Rangan Majumder}, \bibinfo{person}{Andrew McNamara},
  \bibinfo{person}{Bhaskar Mitra}, {and} \bibinfo{person}{Tri Nguyen}.}
  \bibinfo{year}{2016}\natexlab{}.
\newblock \showarticletitle{{MS MARCO : A Human Generated MAchine Reading
  COmprehension Dataset}}. In \bibinfo{booktitle}{\emph{Proc. of NIPS}}.
\newblock


\bibitem[\protect\citeauthoryear{Cao, Qin, Liu, Tsai, and Li}{Cao
  et~al\mbox{.}}{2007}]%
        {cao2007learning}
\bibfield{author}{\bibinfo{person}{Zhe Cao}, \bibinfo{person}{Tao Qin},
  \bibinfo{person}{Tie-Yan Liu}, \bibinfo{person}{Ming-Feng Tsai}, {and}
  \bibinfo{person}{Hang Li}.} \bibinfo{year}{2007}\natexlab{}.
\newblock \showarticletitle{Learning to Rank: From Pairwise Approach to
  Listwise Approach}. In \bibinfo{booktitle}{\emph{Proc. of ICML}}.
\newblock


\bibitem[\protect\citeauthoryear{Caron, Misra, Mairal, Goyal, Bojanowski, and
  Joulin}{Caron et~al\mbox{.}}{2021}]%
        {caron2021unsupervised}
\bibfield{author}{\bibinfo{person}{Mathilde Caron}, \bibinfo{person}{Ishan
  Misra}, \bibinfo{person}{Julien Mairal}, \bibinfo{person}{Priya Goyal},
  \bibinfo{person}{Piotr Bojanowski}, {and} \bibinfo{person}{Armand Joulin}.}
  \bibinfo{year}{2021}\natexlab{}.
\newblock \showarticletitle{Unsupervised Learning of Visual Features by
  Contrasting Cluster Assignments}.
\newblock \bibinfo{journal}{\emph{arXiv:2006.09882}} (\bibinfo{year}{2021}).
\newblock


\bibitem[\protect\citeauthoryear{Chen, He, Hui, Sun, and Sun}{Chen
  et~al\mbox{.}}{2020}]%
        {chen2020simplified}
\bibfield{author}{\bibinfo{person}{Xuanang Chen}, \bibinfo{person}{Ben He},
  \bibinfo{person}{Kai Hui}, \bibinfo{person}{Le Sun}, {and}
  \bibinfo{person}{Yingfei Sun}.} \bibinfo{year}{2020}\natexlab{}.
\newblock \showarticletitle{Simplified {TinyBERT}: Knowledge Distillation for
  Document Retrieval}.
\newblock \bibinfo{journal}{\emph{arXiv:2009.07531}} (\bibinfo{year}{2020}).
\newblock


\bibitem[\protect\citeauthoryear{Cohen, Jordan, and Croft}{Cohen
  et~al\mbox{.}}{2019}]%
        {cohen2019}
\bibfield{author}{\bibinfo{person}{Daniel Cohen}, \bibinfo{person}{Scott~M.
  Jordan}, {and} \bibinfo{person}{W.~Bruce Croft}.}
  \bibinfo{year}{2019}\natexlab{}.
\newblock \showarticletitle{Learning a Better Negative Sampling Policy with
  Deep Neural Networks for Search}. In \bibinfo{booktitle}{\emph{Proc. of
  ICTIR}}.
\newblock


\bibitem[\protect\citeauthoryear{Craswell, Mitra, Yilmaz, and Campos}{Craswell
  et~al\mbox{.}}{2019}]%
        {trec2019overview}
\bibfield{author}{\bibinfo{person}{Nick Craswell}, \bibinfo{person}{Bhaskar
  Mitra}, \bibinfo{person}{Emine Yilmaz}, {and} \bibinfo{person}{Daniel
  Campos}.} \bibinfo{year}{2019}\natexlab{}.
\newblock \showarticletitle{Overview of the TREC 2019 Deep Learning Track}. In
  \bibinfo{booktitle}{\emph{TREC}}.
\newblock


\bibitem[\protect\citeauthoryear{Craswell, Mitra, Yilmaz, and Campos}{Craswell
  et~al\mbox{.}}{2020}]%
        {trec2020overview}
\bibfield{author}{\bibinfo{person}{Nick Craswell}, \bibinfo{person}{Bhaskar
  Mitra}, \bibinfo{person}{Emine Yilmaz}, {and} \bibinfo{person}{Daniel
  Campos}.} \bibinfo{year}{2020}\natexlab{}.
\newblock \showarticletitle{Overview of the TREC 2020 Deep Learning Track}. In
  \bibinfo{booktitle}{\emph{TREC}}.
\newblock


\bibitem[\protect\citeauthoryear{Dai and Callan}{Dai and Callan}{2019}]%
        {dai2019context}
\bibfield{author}{\bibinfo{person}{Zhuyun Dai} {and} \bibinfo{person}{Jamie
  Callan}.} \bibinfo{year}{2019}\natexlab{}.
\newblock \showarticletitle{Context-Aware Sentence/Passage Term Importance
  Estimation for First Stage Retrieval}.
\newblock \bibinfo{journal}{\emph{arXiv:1910.10687}} (\bibinfo{year}{2019}).
\newblock


\bibitem[\protect\citeauthoryear{Devlin, Chang, Lee, and Toutanova}{Devlin
  et~al\mbox{.}}{2019}]%
        {devlin2018bert}
\bibfield{author}{\bibinfo{person}{Jacob Devlin}, \bibinfo{person}{Ming-Wei
  Chang}, \bibinfo{person}{Kenton Lee}, {and} \bibinfo{person}{Kristina
  Toutanova}.} \bibinfo{year}{2019}\natexlab{}.
\newblock \showarticletitle{{BERT}: Pre-training of Deep Bidirectional
  Transformers for Language Understanding}. In \bibinfo{booktitle}{\emph{Proc.
  of NAACL}}.
\newblock


\bibitem[\protect\citeauthoryear{Ding, Liu, Liu, Ren, Zhao, Dong, Wu, and
  Wang}{Ding et~al\mbox{.}}{2020}]%
        {ding2020rocketqa}
\bibfield{author}{\bibinfo{person}{Yingqi Qu~Yuchen Ding},
  \bibinfo{person}{Jing Liu}, \bibinfo{person}{Kai Liu},
  \bibinfo{person}{Ruiyang Ren}, \bibinfo{person}{Xin Zhao},
  \bibinfo{person}{Daxiang Dong}, \bibinfo{person}{Hua Wu}, {and}
  \bibinfo{person}{Haifeng Wang}.} \bibinfo{year}{2020}\natexlab{}.
\newblock \showarticletitle{RocketQA: An Optimized Training Approach to Dense
  Passage Retrieval for Open-Domain Question Answering}.
\newblock \bibinfo{journal}{\emph{arXiv:2010.08191}} (\bibinfo{year}{2020}).
\newblock


\bibitem[\protect\citeauthoryear{Gao, Dai, and Callan}{Gao
  et~al\mbox{.}}{2020}]%
        {gao2020understanding}
\bibfield{author}{\bibinfo{person}{Luyu Gao}, \bibinfo{person}{Zhuyun Dai},
  {and} \bibinfo{person}{Jamie Callan}.} \bibinfo{year}{2020}\natexlab{}.
\newblock \showarticletitle{Understanding BERT Rankers Under Distillation}.
\newblock \bibinfo{journal}{\emph{arXiv:2007.11088}} (\bibinfo{year}{2020}).
\newblock


\bibitem[\protect\citeauthoryear{Gao, Dai, and Callan}{Gao
  et~al\mbox{.}}{2021}]%
        {gao2021rethink}
\bibfield{author}{\bibinfo{person}{Luyu Gao}, \bibinfo{person}{Zhuyun Dai},
  {and} \bibinfo{person}{Jamie Callan}.} \bibinfo{year}{2021}\natexlab{}.
\newblock \showarticletitle{Rethink Training of BERT Rerankers in Multi-Stage
  Retrieval Pipeline}.
\newblock \bibinfo{journal}{\emph{arXiv:2101.08751}} (\bibinfo{year}{2021}).
\newblock


\bibitem[\protect\citeauthoryear{Hofst{\"a}tter, Althammer, Schr{\"o}der,
  Sertkan, and Hanbury}{Hofst{\"a}tter et~al\mbox{.}}{2020}]%
        {hofstaetter2020_crossarchitecture_kd}
\bibfield{author}{\bibinfo{person}{Sebastian Hofst{\"a}tter},
  \bibinfo{person}{Sophia Althammer}, \bibinfo{person}{Michael Schr{\"o}der},
  \bibinfo{person}{Mete Sertkan}, {and} \bibinfo{person}{Allan Hanbury}.}
  \bibinfo{year}{2020}\natexlab{}.
\newblock \showarticletitle{Improving Efficient Neural Ranking Models with
  Cross-Architecture Knowledge Distillation}.
\newblock \bibinfo{journal}{\emph{arXiv:2010.02666}} (\bibinfo{year}{2020}).
\newblock


\bibitem[\protect\citeauthoryear{Hofst{\"a}tter and Hanbury}{Hofst{\"a}tter and
  Hanbury}{2019}]%
        {Hofstaetter2019_osirrc}
\bibfield{author}{\bibinfo{person}{Sebastian Hofst{\"a}tter} {and}
  \bibinfo{person}{Allan Hanbury}.} \bibinfo{year}{2019}\natexlab{}.
\newblock \showarticletitle{Let's Measure Run Time! Extending the IR
  Replicability Infrastructure to Include Performance Aspects}. In
  \bibinfo{booktitle}{\emph{Proc. of OSIRRC}}.
\newblock


\bibitem[\protect\citeauthoryear{Izacard and Grave}{Izacard and Grave}{2020}]%
        {izacard2020distilling}
\bibfield{author}{\bibinfo{person}{Gautier Izacard} {and}
  \bibinfo{person}{Edouard Grave}.} \bibinfo{year}{2020}\natexlab{}.
\newblock \showarticletitle{Distilling Knowledge from Reader to Retriever for
  Question Answering}.
\newblock \bibinfo{journal}{\emph{arXiv:2012.04584}} (\bibinfo{year}{2020}).
\newblock


\bibitem[\protect\citeauthoryear{Jardine and van Rijsbergen}{Jardine and van
  Rijsbergen}{1971}]%
        {jardine1971use}
\bibfield{author}{\bibinfo{person}{Nick Jardine} {and}
  \bibinfo{person}{Cornelis~Joost van Rijsbergen}.}
  \bibinfo{year}{1971}\natexlab{}.
\newblock \showarticletitle{The Use of Hierarchic Clustering in Information
  Retrieval}.
\newblock \bibinfo{journal}{\emph{Information Storage and Retrieval}}
  \bibinfo{volume}{7}, \bibinfo{number}{5} (\bibinfo{year}{1971}),
  \bibinfo{pages}{217--240}.
\newblock


\bibitem[\protect\citeauthoryear{Jiao, Yin, Shang, Jiang, Chen, Li, Wang, and
  Liu}{Jiao et~al\mbox{.}}{2019}]%
        {jiao2019tinybert}
\bibfield{author}{\bibinfo{person}{Xiaoqi Jiao}, \bibinfo{person}{Yichun Yin},
  \bibinfo{person}{Lifeng Shang}, \bibinfo{person}{Xin Jiang},
  \bibinfo{person}{Xiao Chen}, \bibinfo{person}{Linlin Li},
  \bibinfo{person}{Fang Wang}, {and} \bibinfo{person}{Qun Liu}.}
  \bibinfo{year}{2019}\natexlab{}.
\newblock \showarticletitle{{TinyBERT}: Distilling {BERT} for Natural Language
  Understanding}.
\newblock \bibinfo{journal}{\emph{arXiv:1909.10351}} (\bibinfo{year}{2019}).
\newblock


\bibitem[\protect\citeauthoryear{Johnson, Douze, and J{\'e}gou}{Johnson
  et~al\mbox{.}}{2017}]%
        {faiss2017}
\bibfield{author}{\bibinfo{person}{Jeff Johnson}, \bibinfo{person}{Matthijs
  Douze}, {and} \bibinfo{person}{Herv{\'e} J{\'e}gou}.}
  \bibinfo{year}{2017}\natexlab{}.
\newblock \showarticletitle{Billion-Scale Similarity Search with {GPUs}}.
\newblock \bibinfo{journal}{\emph{arXiv:1702.08734}} (\bibinfo{year}{2017}).
\newblock


\bibitem[\protect\citeauthoryear{Kalantidis, Sariyildiz, Pion, Weinzaepfel, and
  Larlus}{Kalantidis et~al\mbox{.}}{2020}]%
        {kalantidis2020hard}
\bibfield{author}{\bibinfo{person}{Yannis Kalantidis},
  \bibinfo{person}{Mert~Bulent Sariyildiz}, \bibinfo{person}{Noe Pion},
  \bibinfo{person}{Philippe Weinzaepfel}, {and} \bibinfo{person}{Diane
  Larlus}.} \bibinfo{year}{2020}\natexlab{}.
\newblock \showarticletitle{Hard Negative Mixing for Contrastive Learning}.
\newblock \bibinfo{journal}{\emph{Advances in Neural Information Processing
  Systems}}  \bibinfo{volume}{33} (\bibinfo{year}{2020}).
\newblock


\bibitem[\protect\citeauthoryear{Karpukhin, Oğuz, Min, Lewis, Wu, Edunov,
  Chen, and tau Yih}{Karpukhin et~al\mbox{.}}{2020}]%
        {karpukhin2020dense}
\bibfield{author}{\bibinfo{person}{Vladimir Karpukhin}, \bibinfo{person}{Barlas
  Oğuz}, \bibinfo{person}{Sewon Min}, \bibinfo{person}{Patrick Lewis},
  \bibinfo{person}{Ledell Wu}, \bibinfo{person}{Sergey Edunov},
  \bibinfo{person}{Danqi Chen}, {and} \bibinfo{person}{Wen tau Yih}.}
  \bibinfo{year}{2020}\natexlab{}.
\newblock \showarticletitle{Dense Passage Retrieval for Open-Domain Question
  Answering}.
\newblock \bibinfo{journal}{\emph{arXiv:2004.04906}} (\bibinfo{year}{2020}).
\newblock


\bibitem[\protect\citeauthoryear{Khattab and Zaharia}{Khattab and
  Zaharia}{2020}]%
        {khattab2020colbert}
\bibfield{author}{\bibinfo{person}{Omar Khattab} {and} \bibinfo{person}{Matei
  Zaharia}.} \bibinfo{year}{2020}\natexlab{}.
\newblock \showarticletitle{ColBERT: Efficient and Effective Passage Search via
  Contextualized Late Interaction over BERT}. In
  \bibinfo{booktitle}{\emph{Proc. of SIGIR}}.
\newblock


\bibitem[\protect\citeauthoryear{Kingma and Ba}{Kingma and Ba}{2014}]%
        {kingma2014adam}
\bibfield{author}{\bibinfo{person}{Diederik~P. Kingma} {and}
  \bibinfo{person}{Jimmy Ba}.} \bibinfo{year}{2014}\natexlab{}.
\newblock \showarticletitle{Adam: A Method for Stochastic Optimization}.
\newblock \bibinfo{journal}{\emph{arXiv:1412.6980}} (\bibinfo{year}{2014}).
\newblock


\bibitem[\protect\citeauthoryear{Lin, Yang, and Lin}{Lin et~al\mbox{.}}{2020}]%
        {lin2020distilling}
\bibfield{author}{\bibinfo{person}{Sheng-Chieh Lin},
  \bibinfo{person}{Jheng-Hong Yang}, {and} \bibinfo{person}{Jimmy Lin}.}
  \bibinfo{year}{2020}\natexlab{}.
\newblock \showarticletitle{Distilling Dense Representations for Ranking using
  Tightly-Coupled Teachers}.
\newblock \bibinfo{journal}{\emph{arXiv:2010.11386}} (\bibinfo{year}{2020}).
\newblock


\bibitem[\protect\citeauthoryear{Lu, Jiao, and Zhang}{Lu et~al\mbox{.}}{2020}]%
        {lu2020twinbert}
\bibfield{author}{\bibinfo{person}{Wenhao Lu}, \bibinfo{person}{Jian Jiao},
  {and} \bibinfo{person}{Ruofei Zhang}.} \bibinfo{year}{2020}\natexlab{}.
\newblock \showarticletitle{TwinBERT: Distilling Knowledge to Twin-Structured
  BERT Models for Efficient Retrieval}.
\newblock \bibinfo{journal}{\emph{arXiv:2002.06275}} (\bibinfo{year}{2020}).
\newblock


\bibitem[\protect\citeauthoryear{Luan, Eisenstein, Toutanova, and Collins}{Luan
  et~al\mbox{.}}{2020}]%
        {luan2020sparse}
\bibfield{author}{\bibinfo{person}{Yi Luan}, \bibinfo{person}{Jacob
  Eisenstein}, \bibinfo{person}{Kristina Toutanova}, {and}
  \bibinfo{person}{Michael Collins}.} \bibinfo{year}{2020}\natexlab{}.
\newblock \showarticletitle{Sparse, Dense, and Attentional Representations for
  Text Retrieval}.
\newblock \bibinfo{journal}{\emph{arXiv:2005.00181}} (\bibinfo{year}{2020}).
\newblock


\bibitem[\protect\citeauthoryear{MacAvaney, Nardini, Perego, Tonellotto,
  Goharian, and Frieder}{MacAvaney et~al\mbox{.}}{2020}]%
        {macavaney2020training}
\bibfield{author}{\bibinfo{person}{Sean MacAvaney},
  \bibinfo{person}{Franco~Maria Nardini}, \bibinfo{person}{Raffaele Perego},
  \bibinfo{person}{Nicola Tonellotto}, \bibinfo{person}{Nazli Goharian}, {and}
  \bibinfo{person}{Ophir Frieder}.} \bibinfo{year}{2020}\natexlab{}.
\newblock \showarticletitle{Training Curricula for Open Domain Answer
  Re-Ranking}. In \bibinfo{booktitle}{\emph{Proc. of SIGIR}}.
\newblock


\bibitem[\protect\citeauthoryear{MacAvaney, Yates, Cohan, and
  Goharian}{MacAvaney et~al\mbox{.}}{2019}]%
        {macavaney2019}
\bibfield{author}{\bibinfo{person}{Sean MacAvaney}, \bibinfo{person}{Andrew
  Yates}, \bibinfo{person}{Arman Cohan}, {and} \bibinfo{person}{Nazli
  Goharian}.} \bibinfo{year}{2019}\natexlab{}.
\newblock \showarticletitle{CEDR: Contextualized Embeddings for Document
  Ranking}. In \bibinfo{booktitle}{\emph{Proc. of SIGIR}}.
\newblock


\bibitem[\protect\citeauthoryear{MacQueen}{MacQueen}{1967}]%
        {macqueen1967some}
\bibfield{author}{\bibinfo{person}{James MacQueen}.}
  \bibinfo{year}{1967}\natexlab{}.
\newblock \showarticletitle{Some Methods for Classification and Analysis of
  Multivariate observations}. In \bibinfo{booktitle}{\emph{Proc. of the Fifth
  Berkeley Symposium on Mathematical Statistics and Probability}},
  Vol.~\bibinfo{volume}{1}. \bibinfo{pages}{281--297}.
\newblock


\bibitem[\protect\citeauthoryear{Nogueira and Cho}{Nogueira and Cho}{2019}]%
        {nogueira2019passage}
\bibfield{author}{\bibinfo{person}{Rodrigo Nogueira} {and}
  \bibinfo{person}{Kyunghyun Cho}.} \bibinfo{year}{2019}\natexlab{}.
\newblock \showarticletitle{Passage Re-ranking with BERT}.
\newblock \bibinfo{journal}{\emph{arXiv:1901.04085}} (\bibinfo{year}{2019}).
\newblock


\bibitem[\protect\citeauthoryear{Nogueira and Lin}{Nogueira and Lin}{2019}]%
        {nogueira2019doct5query}
\bibfield{author}{\bibinfo{person}{Rodrigo Nogueira} {and}
  \bibinfo{person}{Jimmy Lin}.} \bibinfo{year}{2019}\natexlab{}.
\newblock \showarticletitle{From doc2query to docTTTTTquery}.
\newblock \bibinfo{journal}{\emph{Online preprint}} (\bibinfo{year}{2019}).
\newblock


\bibitem[\protect\citeauthoryear{Paszke, Gross, Chintala, Chanan, Yang, DeVito,
  Lin, Desmaison, Antiga, and Lerer}{Paszke et~al\mbox{.}}{2017}]%
        {pytorch2017}
\bibfield{author}{\bibinfo{person}{Adam Paszke}, \bibinfo{person}{Sam Gross},
  \bibinfo{person}{Soumith Chintala}, \bibinfo{person}{Gregory Chanan},
  \bibinfo{person}{Edward Yang}, \bibinfo{person}{Zachary DeVito},
  \bibinfo{person}{Zeming Lin}, \bibinfo{person}{Alban Desmaison},
  \bibinfo{person}{Luca Antiga}, {and} \bibinfo{person}{Adam Lerer}.}
  \bibinfo{year}{2017}\natexlab{}.
\newblock \showarticletitle{Automatic Differentiation in PyTorch}. In
  \bibinfo{booktitle}{\emph{Proc. of NIPS-W}}.
\newblock


\bibitem[\protect\citeauthoryear{Pradeep, Nogueira, and Lin}{Pradeep
  et~al\mbox{.}}{2021}]%
        {pradeep2021expando}
\bibfield{author}{\bibinfo{person}{Ronak Pradeep}, \bibinfo{person}{Rodrigo
  Nogueira}, {and} \bibinfo{person}{Jimmy Lin}.}
  \bibinfo{year}{2021}\natexlab{}.
\newblock \showarticletitle{The Expando-Mono-Duo Design Pattern for Text
  Ranking with Pretrained Sequence-to-Sequence Models}.
\newblock \bibinfo{journal}{\emph{arXiv:2101.05667}} (\bibinfo{year}{2021}).
\newblock


\bibitem[\protect\citeauthoryear{Sanh, Debut, Chaumond, and Wolf}{Sanh
  et~al\mbox{.}}{2019}]%
        {sanh2019distilbert}
\bibfield{author}{\bibinfo{person}{Victor Sanh}, \bibinfo{person}{Lysandre
  Debut}, \bibinfo{person}{Julien Chaumond}, {and} \bibinfo{person}{Thomas
  Wolf}.} \bibinfo{year}{2019}\natexlab{}.
\newblock \showarticletitle{DistilBERT, A Distilled Version of BERT: Smaller,
  Faster, Cheaper and Lighter}.
\newblock \bibinfo{journal}{\emph{arXiv:1910.01108}} (\bibinfo{year}{2019}).
\newblock


\bibitem[\protect\citeauthoryear{Shen, Liu, Liu, Savvides, and Darrell}{Shen
  et~al\mbox{.}}{2020}]%
        {shen2020rethinking}
\bibfield{author}{\bibinfo{person}{Zhiqiang Shen}, \bibinfo{person}{Zechun
  Liu}, \bibinfo{person}{Zhuang Liu}, \bibinfo{person}{Marios Savvides}, {and}
  \bibinfo{person}{Trevor Darrell}.} \bibinfo{year}{2020}\natexlab{}.
\newblock \showarticletitle{Rethinking Image Mixture for Unsupervised Visual
  Representation Learning}.
\newblock \bibinfo{journal}{\emph{arXiv:2003.05438}} (\bibinfo{year}{2020}).
\newblock


\bibitem[\protect\citeauthoryear{Tang and Wang}{Tang and Wang}{2018}]%
        {tang2018ranking}
\bibfield{author}{\bibinfo{person}{Jiaxi Tang} {and} \bibinfo{person}{Ke
  Wang}.} \bibinfo{year}{2018}\natexlab{}.
\newblock \showarticletitle{Ranking Distillation: Learning Compact Ranking
  Models with High Performance for Recommender System}. In
  \bibinfo{booktitle}{\emph{Proc. of SIGKDD}}.
\newblock


\bibitem[\protect\citeauthoryear{Vakili~Tahami, Ghajar, and
  Shakery}{Vakili~Tahami et~al\mbox{.}}{2020}]%
        {vakili2020distilling}
\bibfield{author}{\bibinfo{person}{Amir Vakili~Tahami}, \bibinfo{person}{Kamyar
  Ghajar}, {and} \bibinfo{person}{Azadeh Shakery}.}
  \bibinfo{year}{2020}\natexlab{}.
\newblock \showarticletitle{Distilling Knowledge for Fast Retrieval-based
  Chat-bots}. In \bibinfo{booktitle}{\emph{Proc. of SIGIR}}.
\newblock


\bibitem[\protect\citeauthoryear{Vaswani, Shazeer, Parmar, Uszkoreit, Jones,
  Gomez, Kaiser, and Polosukhin}{Vaswani et~al\mbox{.}}{2017}]%
        {vaswani2017attention}
\bibfield{author}{\bibinfo{person}{Ashish Vaswani}, \bibinfo{person}{Noam
  Shazeer}, \bibinfo{person}{Niki Parmar}, \bibinfo{person}{Jakob Uszkoreit},
  \bibinfo{person}{Llion Jones}, \bibinfo{person}{Aidan~N. Gomez},
  \bibinfo{person}{{\L}ukasz Kaiser}, {and} \bibinfo{person}{Illia
  Polosukhin}.} \bibinfo{year}{2017}\natexlab{}.
\newblock \showarticletitle{Attention Is All You Need}. In
  \bibinfo{booktitle}{\emph{Proc. of NIPS}}.
\newblock


\bibitem[\protect\citeauthoryear{Voorhees}{Voorhees}{1985}]%
        {voorhees1985cluster}
\bibfield{author}{\bibinfo{person}{Ellen~M. Voorhees}.}
  \bibinfo{year}{1985}\natexlab{}.
\newblock \showarticletitle{The Cluster Hypothesis Revisited}. In
  \bibinfo{booktitle}{\emph{Proc. of SIGIR}}.
\newblock


\bibitem[\protect\citeauthoryear{Wang, Li, Golbandi, Bendersky, and
  Najork}{Wang et~al\mbox{.}}{2018}]%
        {wang2018lambdaloss}
\bibfield{author}{\bibinfo{person}{Xuanhui Wang}, \bibinfo{person}{Cheng Li},
  \bibinfo{person}{Nadav Golbandi}, \bibinfo{person}{Michael Bendersky}, {and}
  \bibinfo{person}{Marc Najork}.} \bibinfo{year}{2018}\natexlab{}.
\newblock \showarticletitle{The LambdaLoss Framework for Ranking Metric
  Optimization}. In \bibinfo{booktitle}{\emph{Proc. of CIKM}}.
\newblock


\bibitem[\protect\citeauthoryear{Wolf, Debut, Sanh, Chaumond, Delangue, Moi,
  Cistac, Rault, Louf, Funtowicz, Davison, Shleifer, von Platen, Ma, Jernite,
  Plu, Xu, Le~Scao, Gugger, Drame, Lhoest, and Rush}{Wolf
  et~al\mbox{.}}{2020}]%
        {wolf2019huggingface}
\bibfield{author}{\bibinfo{person}{Thomas Wolf}, \bibinfo{person}{Lysandre
  Debut}, \bibinfo{person}{Victor Sanh}, \bibinfo{person}{Julien Chaumond},
  \bibinfo{person}{Clement Delangue}, \bibinfo{person}{Anthony Moi},
  \bibinfo{person}{Pierric Cistac}, \bibinfo{person}{Tim Rault},
  \bibinfo{person}{Remi Louf}, \bibinfo{person}{Morgan Funtowicz},
  \bibinfo{person}{Joe Davison}, \bibinfo{person}{Sam Shleifer},
  \bibinfo{person}{Patrick von Platen}, \bibinfo{person}{Clara Ma},
  \bibinfo{person}{Yacine Jernite}, \bibinfo{person}{Julien Plu},
  \bibinfo{person}{Canwen Xu}, \bibinfo{person}{Teven Le~Scao},
  \bibinfo{person}{Sylvain Gugger}, \bibinfo{person}{Mariama Drame},
  \bibinfo{person}{Quentin Lhoest}, {and} \bibinfo{person}{Alexander Rush}.}
  \bibinfo{year}{2020}\natexlab{}.
\newblock \showarticletitle{Transformers: State-of-the-Art Natural Language
  Processing}. In \bibinfo{booktitle}{\emph{Proc. EMNLP: System
  Demonstrations}}. \bibinfo{pages}{38--45}.
\newblock


\bibitem[\protect\citeauthoryear{Xiong, Xiong, Li, Tang, Liu, Bennett, Ahmed,
  and Overwijk}{Xiong et~al\mbox{.}}{2020}]%
        {xiong2020approximate}
\bibfield{author}{\bibinfo{person}{Lee Xiong}, \bibinfo{person}{Chenyan Xiong},
  \bibinfo{person}{Ye Li}, \bibinfo{person}{Kwok-Fung Tang},
  \bibinfo{person}{Jialin Liu}, \bibinfo{person}{Paul Bennett},
  \bibinfo{person}{Junaid Ahmed}, {and} \bibinfo{person}{Arnold Overwijk}.}
  \bibinfo{year}{2020}\natexlab{}.
\newblock \showarticletitle{Approximate Nearest Neighbor Negative Contrastive
  Learning for Dense Text Retrieval}.
\newblock \bibinfo{journal}{\emph{arXiv:2007.00808}} (\bibinfo{year}{2020}).
\newblock


\bibitem[\protect\citeauthoryear{Yang, Fang, and Lin}{Yang
  et~al\mbox{.}}{2017}]%
        {Yang2017}
\bibfield{author}{\bibinfo{person}{Peilin Yang}, \bibinfo{person}{Hui Fang},
  {and} \bibinfo{person}{Jimmy Lin}.} \bibinfo{year}{2017}\natexlab{}.
\newblock \showarticletitle{Anserini: Enabling the Use of {Lucene} for
  Information Retrieval Research}. In \bibinfo{booktitle}{\emph{Proc. of
  SIGIR}}.
\newblock


\bibitem[\protect\citeauthoryear{Zhan, Mao, Liu, Zhang, and Ma}{Zhan
  et~al\mbox{.}}{2020}]%
        {zhan2020learning}
\bibfield{author}{\bibinfo{person}{Jingtao Zhan}, \bibinfo{person}{Jiaxin Mao},
  \bibinfo{person}{Yiqun Liu}, \bibinfo{person}{Min Zhang}, {and}
  \bibinfo{person}{Shaoping Ma}.} \bibinfo{year}{2020}\natexlab{}.
\newblock \showarticletitle{Learning To Retrieve: How to Train a Dense
  Retrieval Model Effectively and Efficiently}.
\newblock \bibinfo{journal}{\emph{arXiv:2010.10469}} (\bibinfo{year}{2020}).
\newblock


\bibitem[\protect\citeauthoryear{Zobel}{Zobel}{1998}]%
        {zobel1998reliable}
\bibfield{author}{\bibinfo{person}{Justin Zobel}.}
  \bibinfo{year}{1998}\natexlab{}.
\newblock \showarticletitle{How Reliable are the Results of Large-Scale
  Information Retrieval Experiments?}. In \bibinfo{booktitle}{\emph{Proc. of
  SIGIR}}.
\newblock


\end{thebibliography}
\end{document}